\documentclass[12pt]{iopart}
\usepackage{iopams,epsfig}

\newcommand{\APJ}{{\it Astrophys.~J.\ }}
\newcommand{\APJS}{{\it Astrophys.~J.\ Suppl.\ }}

\newcommand{\MNRAS}{{\it Mon.\ Not.\ R.~Astron.\ Soc.\ }}
\newcommand{\NF}{{\it Nucl.\ Fusion\ }}

\newcommand{\PF}{{\it Phys.\ Fluids\ }}
\newcommand{\PFB}{{\it Phys.\ Fluids~B\ }}
\newcommand{\PoP}{{\it Phys.\ Plasmas\ }}

\newcommand{\JFM}{{\it J.~Fluid Mech.\ }}
\newcommand{\JCompP}{{\it J.~Comput.\ Phys.\ }}

\newcommand{\eqref}[1]{\eref{#1}}

\newcommand{\eqsand}[2]{\eref{#1} and \eref{#2}}
\newcommand{\Eqref}[1]{Equation \eref{#1}}

\newcommand{\figref}[1]{figure \ref{#1}}
\newcommand{\figsref}[1]{figures \ref{#1}}
\newcommand{\Figref}[1]{Figure \ref{#1}}

\newcommand{\secref}[1]{\S\,\ref{#1}}
\newcommand{\apref}[1]{\ref{#1}} 

\newcommand{\bea}{\begin{eqnarray}}
\newcommand{\eea}{\end{eqnarray}}
\newcommand{\beq}{\begin{equation}}
\newcommand{\eeq}{\end{equation}}
\newcommand{\lt}{\left}
\newcommand{\rt}{\right}
\newcommand{\la}{\langle}
\newcommand{\ra}{\rangle}

\newcommand{\dd}{\partial}
\newcommand{\vdel}{\bnabla}

\newcommand{\ephi}{\varphi}
\newcommand{\eps}{\varepsilon}
\newcommand{\ee}{\epsilon}
\renewcommand{\Re}{\mathrm{Re}}
\renewcommand{\Im}{\mathrm{Im}}
\newcommand{\Zfn}{\mathcal{Z}}

\newcommand{\du}{\delta v_E}
\newcommand{\chiT}{\chi_T} 
\newcommand{\geff}{\gamma_\mathrm{eff}} 

\newcommand{\vr}{\bi{r}}
\newcommand{\vR}{\bi{R}}
\newcommand{\vk}{\bi{k}}
\newcommand{\vu}{\bi{u}}
\newcommand{\vE}{\bi{E}}

\newcommand{\dphi}{\ephi}
\newcommand{\tphi}{\tilde\ephi}
\newcommand{\vB}{\bi{B}}
\newcommand{\vb}{\hat{\bi{b}}}
\newcommand{\vx}{\hat{\bi{x}}}
\newcommand{\vy}{\hat{\bi{y}}}
\newcommand{\vz}{\hat{\bi{z}}}

\newcommand{\avgs}[1]{\la#1\ra_{\vR_s}}
\newcommand{\avgR}[1]{\la#1\ra_{\vR}}
\newcommand{\avgr}[1]{\la#1\ra_{\vr}}

\newcommand{\kpar}{k_\parallel}
\newcommand{\kperp}{k_\perp}

\newcommand{\LB}{L_s}

\newcommand{\LuT}{L_{uT}}
\newcommand{\LT}{L_T}
\newcommand{\Ln}{L_n}
\newcommand{\omn}{\omega_{*}}

\newcommand{\zz}{\bar{\omega}}
\newcommand{\gbar}{\bar{\gamma}}
\newcommand{\zn}{\zz_*}

\newcommand{\zS}{\zz_S}
\newcommand{\eS}{\eta_S}
\newcommand{\tz}{\tilde{\omega}}
\newcommand{\tg}{\tilde{\gamma}}

\newcommand{\nuii}{\nu_{ii}}

\newcommand{\vthi}{v_{\mathrm{th}i}}
\newcommand{\vth}{v_{\mathrm{th}i}}
\newcommand{\rhoi}{\rho_i}
\newcommand{\vv}{\bi{v}}
\newcommand{\vV}{\bi{V}}
\newcommand{\vw}{\bi{w}}

\newcommand{\wperp}{w_\perp}
\newcommand{\wpar}{w_\parallel}
\newcommand{\sigmapar}{\sigma_\parallel}

\newcommand{\tbar}{\Delta t}

\newcommand{\tpvg}{t_0^\mathrm{(PVG)}}

\newcommand{\Nmax}{N_\mathrm{max}}

\begin{document}

\title[Subcritical fluctuations in rotating plasmas]{Suppression of turbulence and 
subcritical fluctuations in differentially rotating gyrokinetic plasmas}

\author{A~A~Schekochihin,$^{1,2}$ E~G~Highcock$^{1-5}$ and S~C~Cowley$^{4,6}$}
\address{$^1$ Rudolf Peierls Centre for Theoretical Physics, University of Oxford, 
1 Keble Rd, Oxford OX1 3NP, UK}
\address{$^2$ Merton College, Oxford OX1 4JD, UK}
\address{$^3$ Magdalen College, Oxford OX1 4AU, UK}
\address{$^4$ Euratom/CCFE Association, Culham Science Centre, Abingdon OX14 3DB, UK}
\address{$^5$ Wolfgang Pauli Institute, University of Vienna, A1090 Vienna, Austria}
\address{$^6$ Blackett Laboratory, Imperial College, London SW7 2AZ, UK}

\ead{a.schekochihin1@physics.ox.ac.uk}

\begin{abstract}
Differential rotation is known to suppress linear instabilities in fusion plasmas. 
However, numerical experiments show that even in the absence of growing eigenmodes, 
subcritical fluctuations that grow transiently can lead to sustained turbulence, 
limiting the ability of the velocity shear to suppress anomalous transport. 
Here transient growth of electrostatic fluctuations driven by the parallel velocity 
gradient (PVG) and the ion temperature gradient (ITG) in the presence 
of a perpendicular ($\vE\times\vB$) velocity shear is considered. 
The maximally simplified (but most promising for transport reduction) 
case of zero magnetic shear is treated 
in the framework of a local shearing box approximation. In this case 
there are no linearly growing eigenmodes, so all excitations are transient. 
In the PVG-dominated regime, the maximum amplification factor  
is found to be $e^N$ with $N\propto q/\ee$ (safety factor/aspect ratio), the maximally amplified 
wavenumbers perpendicular and parallel to the magnetic field are related by 
$k_y\rhoi\approx (\ee/q)^{1/3}\kpar\vthi/S$, where $\rhoi$ is the ion 
Larmor radius, $\vthi$ the ion thermal speed and $S$ the $\vE\times\vB$ shear.
In the ITG-dominated regime, $N$ is independent of wavenumber and  
$N\propto\vthi/(\LT S)$, where $\LT$ is the ion-temperature scale length. 
Intermediate ITG-PVG regimes are also analysed and $N$ is calculated as a 
function of $q/\ee$, $\LT$ and $S$. Analytical results are 
corroborated and supplemented by linear gyrokinetic numerical tests. 
Regimes with $N\lesssim1$ for all wavenumbers are possible for sufficiently 
low values of $q/\ee$ ($\lesssim7$ in our model); 
ion-scale turbulence is expected to be fully suppressed in such regimes.  
For cases when it is not suppressed, an elementary heuristic theory of subcritical PVG turbulence 
leading to a scaling of the associated ion heat flux with $q$, $\ee$, $S$ and $\LT$ 
is proposed; it is argued that the transport is much less ``stiff'' than in the ITG regime.  
\end{abstract}


\section{Introduction}
\label{sec:intro}

It has long been understood that anomalous transport 
in tokamaks is caused by turbulence at or just above the ion Larmor scale, 
which is powered by drift instabilities extracting free energy from background gradients. 
The ion-temperature-gradient (ITG) instability \cite{Rudakov61,Coppi67,Linsker81} 
and the resulting turbulence \cite{Cowley91} have been identified as 
a particular culprit. With the advent first of gyrofluid and then 
gyrokinetic numerical simulations, ITG turbulence has been the main focus 
of numerical studies aiming to map out levels of turbulent transport 
expected for any given set of equilibrium parameters (mean profile gradients 
and the magnetic configuration) \cite{Kotschenreuther95,Dimits00,Kinsey06,Barnes11itg}. 
These studies assumed that the differential (toroidal) rotation of 
the tokamak plasmas, caused by momentum injection from the heating beams, 
had a negligible effect on the turbulence and on the resulting transport. 
This, however, is known to be an inadequate approximation for many 
devices and configurations in which the rotational shear can be comparable 
to the typical turbulent rate of strain. Linear eigenmode analysis 
\cite{Artun92,Artun93,Dong93,Connor07} suggests that the ITG growth 
rates are reduced and can be completely quenched by $\vE\times\vB$ shear. 
A reduction or even complete suppression of associated transport 
was therefore expected and indeed found numerically, at least 
for some parameter regimes \cite{Kinsey05,Roach09,Casson09} 
--- a result that kindled high hopes for controlling turbulence 
with shear and achieving transport bifurcations to steeper temperature 
gradients \cite{Dorland94,Waltz97}.  

\begin{figure}[t]
\centerline{\epsfig{file=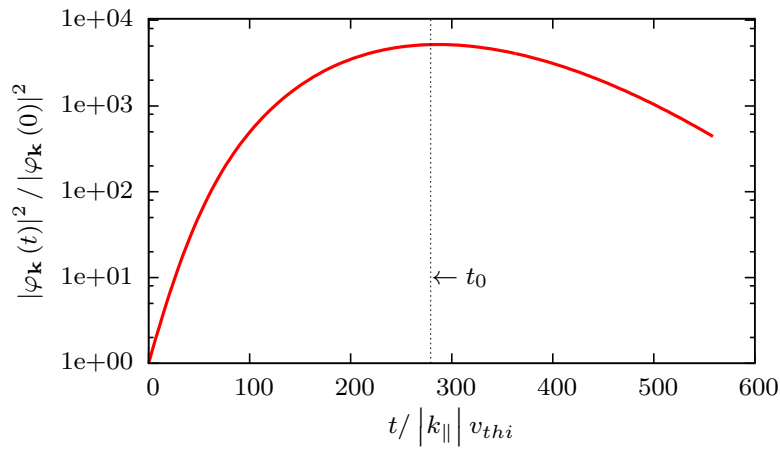,width=4in}}
\caption{\label{fig:trgrowth} 
Time evolution of the squared amplitude (normalised to its initial value) 
of a pure-PVG-driven linear perturbation, 
obtained in a direct linear numerical simulation using the gyrokinetic 
code {\tt AstroGK} \cite{Numata10} with
$1/\Ln=1/\LT=0$, $q/\ee=50$, $\tau/Z=1$, $k_y\rhoi=1$, $\kpar\vth/S=0.5$. 
The effective growth rate for this case is 
shown in \figref{fig:gamma}.} 
\end{figure}

Since the tokamak rotation is (to lowest order in the gyrokinetic expansion) 
purely toroidal \cite{Hinton85,Cowley86,Catto87,Connor87,Abel11}, 
strong perpendicular $\vE\times\vB$ shear 
comes at the price of a (stronger by a factor of $q/\ee$) parallel velocity gradient (PVG) 
--- which is itself a source of free energy and so can trigger 
a drift instability \cite{Catto73}, which in turns gives rise 
to turbulence and anomalous transport \cite{Waltz94,Dimits01,Barnes11}. 
This instability is in fact also suppressed by the $\vE\times\vB$ shear, but 
only in the sense that no unstable eigenmodes survive at large 
enough velocity and small enough magnetic shear: 
transient linear amplification is still possible \cite{Waelbroeck92,Waelbroeck94,Newton10}. 
This transient amplification (illustrated in \figref{fig:trgrowth}) can be sufficient 
to give rise to subcritically excited turbulence \cite{Barnes11,Highcock10,Highcock11}. 
It is this turbulence that limits the effectiveness of differential 
rotation in suppressing anomalous transport. While transport bifurcations 
are still possible \cite{Highcock10,Highcock11,Parra11}, finding 
them in the parameter space turns out to be quite a delicate task, 
mostly because the subcritical PVG turbulence is a largely 
unexplored phenomenon. In particular, it is not amenable to the 
usual ``quasilinear'' mixing-length-type arguments because those are 
based on the calculation of linear growth rates --- and it is not 
obvious what should replace them for subcritical turbulence. 

We take the rather obvious view that addressing this problem should 
start from a systematic understanding of 
the linear transient amplification of gyrokinetic fluctuations in 
the presence of velocity shear. The purest example is presented 
by the limit of zero (negligibly low) magnetic shear, which has the twin 
advantages of analytical simplicity and of being most amenable to 
transport bifurcations or at least reduced levels of turbulent transport 
--- as suggested both by numerical experiments \cite{Highcock10,Highcock11} 
and laboratory measurements \cite{Mantica09,Mantica11}. Formally, this regime 
supports no growing eigenmodes at any finite value of velocity shear, so we 
can focus on transient amplification and its dependence on the 
parameters of the problem (the ion temperature gradient, the velocity 
shear, the safety factor $q$ and the device aspect ratio $\ee$) 
without the complications of dealing with the transition to linear eigenmode
stability. This is the kinetic treatment of this problem, 
following in the footsteps of the fluid theory \cite{Newton10} 
(which was done for a finite magnetic shear; note that zero magnetic shear 
is formally a singular limit). 

Since much of the transport modeling for fusion devices relies on 
gyrokinetic simulations, it is important to ascertain that the linear 
behaviour predicted analytically is reproducible 
in direct gyrokinetic simulations with standard numerical codes. 
In what follows, we do this using 
the {\tt AstroGK} code \cite{Numata10}, which, for all practical purposes,  
is the slab version of the widely used fusion code 
{\tt GS2}\footnote{URL: http://gyrokinetics.sourceforge.net} 
(running {\tt GS2} itself in slab mode produces similar results). 
The code solves \eqsand{eq:gk_sh}{eq:quasineut_sh} with nonlinearity 
switched off and a collision operator \cite{Abel08,Barnes09} added to provide 
small-scale regularisation in phase space. These simulations demonstrate 
the degree to which our asymptotic results represent an adequate description 
of realistic parameter regimes and how the transition from the short-time 
to the long-time limit occurs, subject also to (minor) modifications 
of our results by collisions, finite resolution and numerical inaccuracies. 

The rest of the paper is organised as follows. 
In \secref{sec:gk}, we introduce gyrokinetics in a shearing box
--- the governing equations of the minimal model 
we have chosen to treat. In \secref{sec:pure_pvg}, we consider a further 
simplified limit in which the ITG drive is negligible compared with the 
PVG drive and so we can concentrate on the essential properties of the 
latter, namely, work out how the transient amplification time and 
the amplification exponent depend on the velocity shear, $q$ and $\ee$
and which wavenumbers prove most prone to being amplified. 
In \secref{sec:incl_itg}, we generalise these considerations by including 
the ITG. Both analytical and (linear) numerical results are presented 
throughout. 
A qualitative summary of the linear regime and a comparison with the 
fluid limit treated in \cite{Newton10} are given in \secref{sec:qualit}. 
Finally, in \secref{sec:turb}, a criterion for the onset of subcrtitical PVG turbulence 
is proposed, followed by a very crude heuristic theory of this turbulence and 
of the resulting heat transport --- these considerations provide 
a version of the standard mixing-length ``quasilinear'' arguments suitable 
for transiently growing fluctuations 
and a set of scaling predictions in the spirit of \cite{Barnes11itg}.
In particular, we give a semi-quantitative form to the argument that heat 
transport must be much much less ``stiff'' in the presence of velocity shear 
\cite{Mantica09,Mantica11,Highcock10,Highcock11} than in in the standard 
ITG regime because the turbulence is no longer driven by the temperature gradient. 

\section{Gyrokinetics in a shearing box}
\label{sec:gk}

Consider a plasma in a strong mean magnetic field. On the assumption 
of axisymmetry, this field can be expressed in the form 
$\vB = I(\psi)\vdel\phi + \vdel\psi\times\vdel\phi$,
where $\psi$ (magnetic flux) and $I(\psi)$ are two scalar functions determined by solving 
the mean MHD equilibrium equations and $\phi$ is the azimuthal angle with respect 
to which symmetry is assumed. It can be shown (see \cite{Abel11} and references 
therein) that if such a plasma rotates with some velocity ordered in the gyrokinetic 
expansion as similar in size 
to the speed of sound, this mean rotation must be purely azimuthal, 
the same for both species and with an angular velocity $\omega$ that depends on 
the flux label $\psi$ only:
$\vu = \omega(\psi)R^2\vdel\phi$,
where $R$ is the radial coordinate. Thus, each of the nested toroidal flux surfaces 
rotates at its own rate and there is a velocity shear set up by the variation 
of $\omega$ with $\psi$. In this paper, we will be concerned only with the effect 
on the plasma stability of this differential character of the rotation. 
Formally, this effect is isolated by assuming the Mach number $M = u/\vthi$ ($\vthi$ 
is the ion thermal speed) to be moderately low. Mathematically, this can be cast as 
an expansion in $M\ll1$ (subsidiary to the gyrokinetic expansion) where, 
while the mean velocity is ordered subsonic, $u\sim O(M)$, 
its scale length is ordered as $O(1/M)$.   
Under this scheme, effects such the Coriolis or centrifugal 
motion (as well as other, more obscure, ones) 
that scale with the magnitude of the mean velocity 
are negligible, while velocity gradients are retained. 

\begin{figure}[t]
\centerline{\epsfig{file=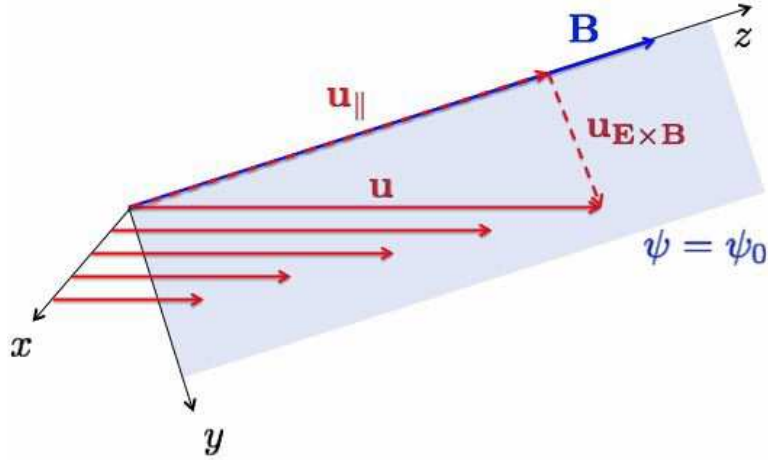,width=4in}}
\caption{\label{fig:xyz} 
Magnetic field, velocity field (sheared) and the local Cartesian frame.
The reference flux surface $\psi=\psi_0$ is shaded.} 
\end{figure}

Let us consider the vicinity of some flux surface $\psi=\psi_0$ and 
introduce a local orthogonal Cartesian frame $(x,y,z)$ that is moving 
with the flux surface $\psi_0$ and in which 
$x$ is the cross-flux-surface (``radial'') coordinate, 
$z$ is the coordinate in the direction of the magnetic field
and $y$ completes the orthogonal frame ($\vy=\vz\times\vx$). Then 
the velocity field can be replaced by a pure linear shear flow:  
\beq
\vu\cdot\vdel\approx Sx\,\frac{\dd}{\dd y},\quad
S = \frac{B_p^2 R^2}{B}\frac{d\omega}{d\psi},
\label{eq:Sdef}
\eeq
where $B_p=|\vdel\psi|/R$ is the poloidal magnetic field (see 
\apref{app:shbox} for details; \figref{fig:xyz} illustrates the geometry). 
This model might be called the ``flying slab approximation,'' 
or, perhaps more conventionally, the shearing box --- a common analytical 
simplification in the fluid dynamics of differentially rotating systems 
\cite{Goldreich65}. 

We will make three further simplifying approximations by assuming 
all fluctuations to be electrostatic, electrons to have purely 
Boltzmann response and the mean magnetic field to be (locally) straight
and uniform, so it has neither curvature nor variation of magnitude nor shear 
(which means that the magnetic drifts can be dropped). 
With all these assumptions and working in the flying slab (shearing box), 
we can write the gyrokinetic system of equations
in the following form (see \apref{app:gk}) 
\bea
\fl
\nonumber
\frac{\dd h}{\dd t} + Sx\frac{\dd h}{\dd y} + \wpar\frac{\dd h}{\dd z} 
+\frac{\rhoi\vth}{2}\lt(\frac{\dd\avgR{\tphi}}{\dd x}\frac{\dd h}{\dd y}
- \frac{\dd\avgR{\tphi}}{\dd y}\frac{\dd h}{\dd x}\rt)
= \lt(\frac{\dd}{\dd t} 
+ Sx\frac{\dd}{\dd y}\rt)\avgR{\tphi}\\
\label{eq:gk_sh}
- \frac{\rhoi\vth}{2}\frac{\dd\avgR{\tphi}}{\dd y}\lt[\frac{1}{\Ln} + 
\lt(\frac{w^2}{\vth^2} - \frac{3}{2}\rt)\frac{1}{\LT} 
-\frac{2\wpar}{\vth^2}\frac{q}{\ee}\,S\rt]F_0,\\
\lt(1+\frac{\tau}{Z}\rt)\tphi = \frac{1}{n_i}\int d^3\vw\avgr{h},
\label{eq:quasineut_sh}
\eea
where $h(t,\vR,w,\wpar)$ is the ion gyrocentre distribution function, 
$F_0(\psi,w)$ the equilibrium Maxwellian (of the ions), 
$\tphi(t,\vr) = Ze\dphi/T_i$ the nondimensionalised 
electrostatic potential ($Ze$ is the ion charge), 
$\rhoi$ the ion Larmor radius, $n_i$ the ion number density, 
$\vw$ the particles' peculiar velocity with respect to the mean flow, 
$\vr$ the position coordinate, $\vR$ the guiding-centre coordinate, 
and the angle brackets denote the gyroaverages holding constant the coordinate 
appearing in their subscript (the precise definitions are provided in \apref{app:gk}). 
We have defined the equilibrium gradients and two other 
standard parameters as follows 
\beq
\frac{1}{\Ln} = -B_pR\frac{d\ln n_i}{d\psi},\quad
\frac{1}{\LT} = -B_pR\frac{d\ln T_i}{d\psi},\quad
\frac{q}{\ee} = \frac{B_\phi}{B_p},\quad
\tau = \frac{T_i}{T_e},
\eeq
where $n_i$ and $T_i$ are the ion density and temperature, respectively, 
and $B_\phi$ is the azimuthal mean magnetic field.  

Note that the velocity shear $S$ appears in two ways in \eqref{eq:gk_sh}: 
as perpendicular shear (multiplying $x\,\dd/\dd y$) and as parallel shear 
(multiplying $F_0$ in the right-hand side). The perpendicular shear rips 
apart the unstable fluctuations and will have a stabilising effect, whereas 
the parallel shear (the PVG) acts as a source of free energy in a manner analogous 
to ITG and drives a drift instability (\secref{sec:pvg}). 
The relative size of these two effects is set by the value of $q/\ee$. 

\subsection{Case of non-zero magnetic shear}

Including a (locally) constant linear magnetic shear into the problem amounts to 
replacing in \eqref{eq:gk_sh}
\beq
Sx\frac{\dd h}{\dd y}\to
\lt(S + \frac{\wpar}{\LB}\rt)x\frac{\dd h}{\dd y},
\eeq
where $\LB$ is the scale length associated with the magnetic shear. 
This appears to introduce complications as we now have an ``effective shear'' 
that depends on the particle velocity $\wpar$. However, since the 
size of $\wpar$ is constrained by the Maxwellian equilibrium distribution, 
this term can be neglected provided $M_s=S\LB/\vth\gg1$. Under this assumption, 
the theory developed below applies without modification. We note
that the ``shear Mach number'' $M_s$ is precisely the parameter 
that is known to control linear stability and transient amplification 
of in the ITG-PVG-driven plasmas in the fluid (collisional) limit 
\cite{Newton10}. 

\subsection{Shearing  frame}

The next step --- standard in treatments of systems with linear shear --- is to make 
a variable transformation $(t,\vr)\to(t',\vr')$ that removes the shear terms ($Sx\dd/\dd y$):
\beq
t'=t,\quad
x'=x,\quad
y'=y-Sxt,\quad
z'=z,
\eeq
and similarly for $(t,\vR)\to(t',\vR')$.
The Fourier transform can then be performed in the primed variables, so
\beq
\tphi = \sum_{\vk'} \tphi_{\vk'}(t') e^{i\vk'\cdot\vr'}
= \sum_{\vk'} \tphi_{\vk'}(t') e^{i\vk(\vk',t')\cdot\vr},
\eeq
where $k_x=k_x'-Sk_y't'$, $k_y=k_y'$, and $\kpar = \kpar'$ (we denote $\kpar\equiv k_z$). 
As usual in the gyrokinetic theory, working in Fourier space allows us to 
compute the gyroaverages in terms of Bessel functions:
\beq
\fl
\avgR{\tphi} = \sum_{\vk'} J_0(a(t'))\tphi_{\vk'}(t') e^{i\vk'\cdot\vR'},
\quad
\avgr{h} = \sum_{\vk'} J_0(a(t')) h_{\vk'}(t') e^{i\vk'\cdot\vr'},
\label{eq:avgs}
\eeq
where $a(t') = \kperp\wperp/\Omega_i = (\wperp/\Omega_i)\sqrt{(k_x'-Sk_y't')^2 + k_y^{\prime 2}}
= (k_y'\wperp/\Omega_i)\sqrt{1 + S^2t^{\prime\prime 2}}$, and we have shifted the origin 
of time: $t''=t'-S^{-1}k_x'/k_y'$. 

Finally, we rewrite the gyrokinetic system \eqref{eq:gk_sh}--\eqref{eq:quasineut_sh} 
in the new variables $(t'',\vk')$. Since we are interested only in the linear problem 
here, we will drop the nonlinearity. We also suppress all primes in the variables.
The result~is
\bea
\fl
\label{eq:gk_lin}
\frac{\dd h_\vk}{\dd t} + i\kpar\wpar h_\vk = \lt\{\frac{\dd}{\dd t}
-i\lt[\omn + \lt(\frac{w^2}{\vth^2}-\frac{3}{2}\rt)\omn\eta_i 
- \frac{\wpar}{\vth}\frac{q}{\ee}\,Sk_y\rhoi\rt]\rt\}F_0J_0(a(t))\tphi_\vk,\\
\lt(1+\frac{\tau}{Z}\rt)\tphi_\vk = \frac{1}{n}\int d^3\vw\,J_0(a(t))h_\vk,
\label{eq:quasineut_lin}
\eea
where $\omn = k_y\rhoi\vth/2\Ln$ is the drift frequency and $\eta_i=\Ln/\LT$; 
the argument of the Bessel function is 
$a(t)=(\wperp/\vth)k_y\rhoi\sqrt{1 + S^2t^2}$. 

\subsection{Integral equation for the linearised problem}
\label{sec:lin_sh}

We integrate \eqref{eq:gk_lin} with respect to time, 
assume the initial fluctuation amplitude small compared to values to which 
it will grow during the subsequent time evolution, 
rescale time  $|\kpar|\vth t \to t$, denote $\tbar=t-t'$, 
use \eqref{eq:quasineut_lin}, in which the velocity integrals  
involving the Maxwellian $F_0=n\,e^{-w^2/\vth^2}/(\pi\vth^2)^{3/2}$ 
are done in the usual way, 
and obtain finally the following integral equation for $\tphi(t)$:
\bea
\nonumber
\fl
\lt(1+\frac{\tau}{Z} - \Gamma_0(\lambda,\lambda)\rt)\tphi(t) 
= \int_0^t d\tbar\,e^{-\tbar^2/4}
\lt\{\lt(\frac{q}{\ee}\,\zS-1\rt)\frac{\tbar}{2}\rt.\\ 
-\lt. i\zn\lt[1 + \eta_i\lt(\Lambda(\lambda,\lambda') - 1 - \frac{\tbar^2}{4}\rt)\rt]\rt\}
\Gamma_0(\lambda,\lambda')\tphi(t-\tbar),
\label{eq:inteq}
\eea
where $\zn=\omn/|\kpar|\vth = k_y\rhoi/2|\kpar|\Ln$ is the normalised 
drift frequency, $\zS = S k_y\rhoi/\kpar\vth$ the normalised shear parameter, and 
\beq
\fl
\Gamma_0(\lambda,\lambda') = e^{-(\lambda+\lambda')/2}I_0(\sqrt{\lambda\lambda'}),\quad
\Lambda(\lambda,\lambda') = 1-\frac{\lambda+\lambda'}{2} + \sqrt{\lambda\lambda'}\,
\frac{I_1(\sqrt{\lambda\lambda'})}{I_0(\sqrt{\lambda\lambda'})}, 
\eeq 
where 
$\lambda(t) = (k_y^2\rhoi^2 + \zS^2t^2)/2$,  
$\lambda'=\lambda(t') = (k_y^2\rhoi^2 + \zS^2t^{\prime 2})/2$, 
and $I_0$ and $I_1$ are modified Bessel functions of the first kind.  

\Eqref{eq:inteq} is the master equation for the linear time evolution of the plasma fluctuations 
driven by the ITG (the $\zn\eta_i$ term) and the PVG (the $(q/\ee)\zS$ term).

\section{Solution for the case of strong shear}
\label{sec:pure_pvg}

We will first consider the maximally simplified case of pure PVG drive (strong shear). 
This is a good quantitative approximation to the general case 
if $\zn, \eta_i\zn \ll (q/\ee)\zS$, which in terms of the basic dimensional 
parameters of the problem translates into 
\beq
\frac{qS}{\ee} \gg \frac{\vth}{\Ln}, \frac{\vth}{\LT}. 
\label{eq:pure_pvg}
\eeq
This is the regime into which the plasma is pushed as the flow shear is increased
--- under certain conditions, the transition can occur abruptly, via 
a transport bifurcation \cite{Highcock10,Parra11,Highcock11}. Besides being, 
therefore, physically the most interesting, this limit 
also has the advantage of particular analytical transparency 
(the more general case including ITG will be considered in \secref{sec:incl_itg}). 

Thus, neglecting all terms that contain $\zn$ and $\eta_i$, \eqref{eq:inteq} becomes
\beq
\fl
\lt(1+\frac{\tau}{Z} - \Gamma_0(\lambda,\lambda)\rt)\tphi(t) 
= \frac{1}{2}\int_0^t d\tbar\,\tbar\,e^{-\tbar^2/4}
\lt(\frac{q}{\ee}\,\zS-1\rt)
\Gamma_0(\lambda,\lambda')\tphi(t-\tbar).
\label{eq:inteq_pvg}
\eeq

\subsection{Short-time limit: the PVG instability}
\label{sec:pvg}

Let us first consider the case in which the velocity shear is unimportant 
except for the PVG drive, i.e., we can approximate 
$\lambda\approx\lambda'\approx k_y^2\rhoi^2/2$ and so there is no time 
dependence in the Bessel functions in \eqref{eq:inteq_pvg}.
We would also like to be able to assume $t\gg1$ 
so that the time integration in \eqref{eq:inteq_pvg} can be extended to $\infty$. 
Formally this limiting case is achieved by ordering $k_y\rhoi\sim1$ and 
$1\ll t \ll 1/\zS\lesssim q/\ee$ 
(in dimensional terms, this is equivalent to $St\ll1$, $\kpar\vth t\gg1$
and $1\ll\kpar\vth/S\lesssim q/\ee$). 
Physically, this regime is realised in the initial stage of evolution 
of the fluctuations or, equivalently, in the case of very weak shear 
but large $q/\ee$.  
 
Under this ordering, we can seek solutions to \eqref{eq:inteq_pvg} 
in the form $\tphi(t) = \tphi_0\exp(-i\zz t)$, where 
$\zz = \omega/|\kpar|\vth$ is the nondimensionalised complex frequency 
and $\omega$ its dimensional counterpart. The time integral in \eqref{eq:inteq_pvg} 
can be expressed in terms of the plasma dispersion function, which 
satisfies \cite{Fried61}
\beq
\label{eq:Z}
\Zfn(\zz) 
= i\int_0^\infty d\tbar\,e^{i\zz\tbar - \tbar^2/4},\qquad
\Zfn'(\zz) = -2[1+\zz\Zfn(\zz)].
\eeq
With the aid of these formulae, \eqref{eq:inteq} is readily converted 
into a transcendental equation for $\zz$: 
\beq
\label{eq:pvg}
\frac{1+\tau/Z}{\Gamma_0(\lambda)} - 1 = 
\lt(\frac{q}{\ee}\,\zS - 1\rt) [1+\zz\Zfn(\zz)], 
\eeq
where $\Gamma_0(\lambda) = e^{-\lambda} I_0(\lambda)$, 
$\lambda=k_y^2\rhoi^2/2$ and $\zS=S k_y\rhoi/\kpar\vth$.


\Eqref{eq:pvg} is simply the dispersion relation for the ion acoustic wave modified 
by the PVG drive term. This point is probably best 
illustrated by considering the cold-ion/long-wavelength limit 
$\tau\ll1$, $\zz=\omega/|\kpar|\vth\gg1$, $\lambda=k_y^2\rho^2/2\ll1$. 
Then $\Gamma_0(\lambda)\approx1$, 
$1+\zz\Zfn(\zz)\approx -1/2\zz^2 + i\zz\sqrt{\pi}\,e^{-\zz^2}$, 
and so, restoring dimensions in \eqref{eq:pvg}, we get
\beq
\zz^2 \approx \frac{Z}{2\tau}\lt(1 - \frac{q}{\ee}\,\zS\rt)
\quad\Rightarrow\quad
\omega \approx \pm\kpar c_s\lt(1 - \frac{qS}{\ee}\frac{k_y\rhoi}{\kpar\vth}\rt)^{1/2},
\label{eq:sound_wave}
\eeq
where in the last expression, we have restored dimensions and denoted 
$c_s=(Z/2\tau)^{1/2}\vth=(T_e/m_i)^{1/2}$, the sound speed.  
When $(q/\ee)\zS$ is sufficiently large, the sound wave is destabilised 
and turns into the PVG instability. Note that it loses its real frequency 
in this transition.

\begin{figure}[t]
\centerline{\epsfig{file=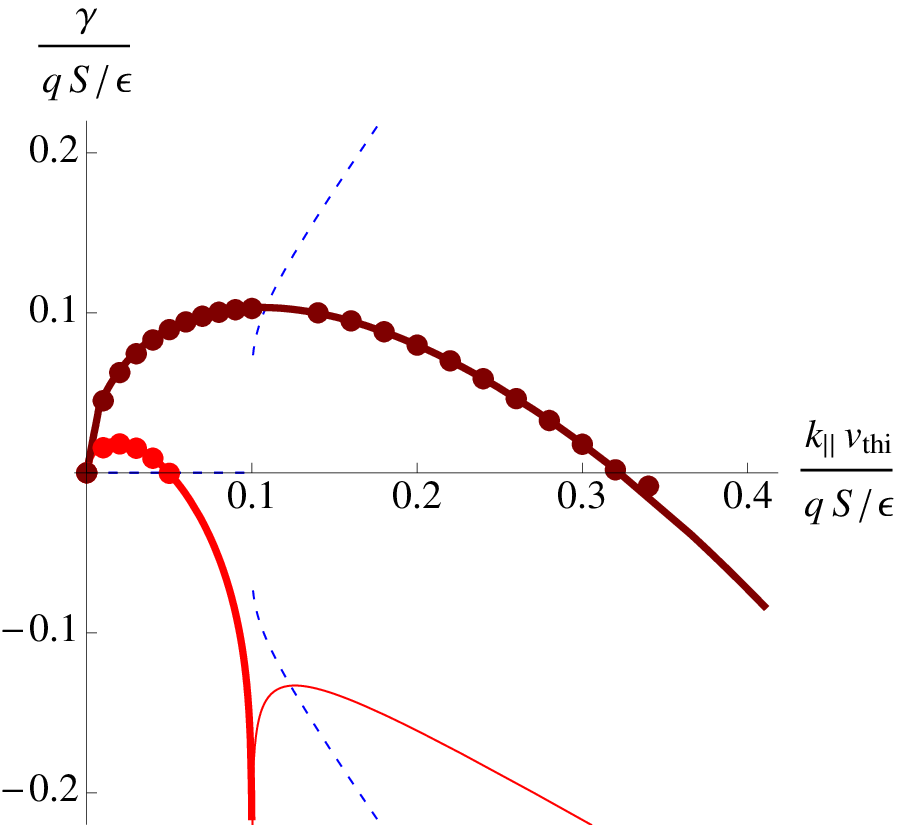,height=2.75in}
\hskip0.1in
\epsfig{file=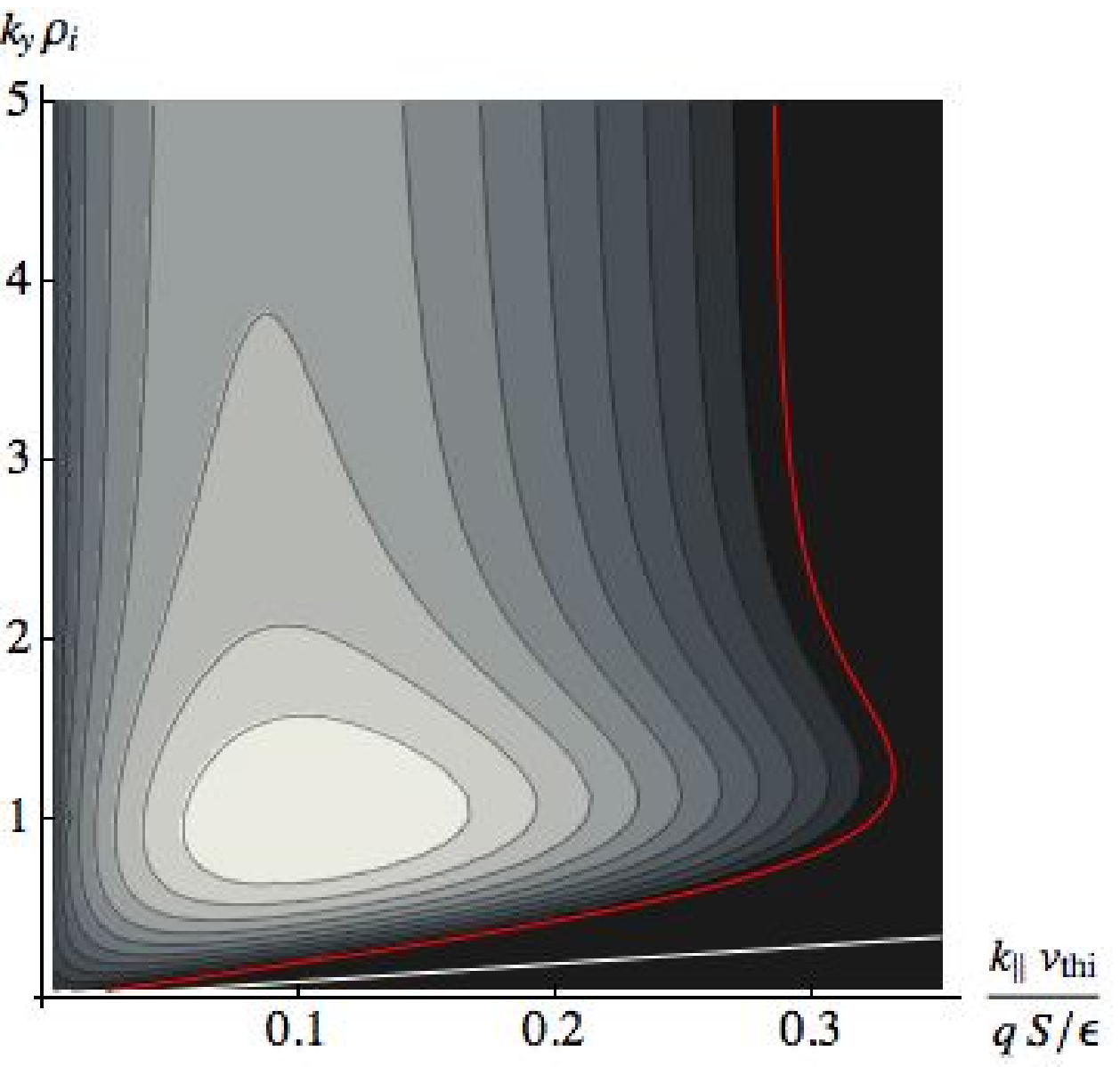,height=2.75in}}
\caption{\label{fig:pvg} {\em Left panel:} 
Growth rate $\gamma = \Im\,\omega$ (normalised by $qS/\ee$) 
vs.\ $\kpar\vth/(qS/\ee)$. Here $\tau/Z=1$, and the two curves 
are for $k_y\rhoi=0.1$ (red) and $k_y\rhoi=1$ (brown).
The growth rate becomes independent of $k_y$ for $k_y\rhoi\gg1$ 
(see end of \apref{app:pvg}) and all curves for large $k_y\rhoi$ are 
very close to each other and similar in shape to $k_y\rhoi=1$ (see right panel). 
The mode has no real frequency for $\kpar\vth/(qS/\ee)<k_y\rhoi$;
for $\kpar\vth/(qS/\ee)>k_y\rhoi$, it turns into a damped sound wave: 
the corresponding frequencies and the damping rate are 
shown only for $k_y\rhoi=0.1$, as thin blue (dashed) and red (solid) lines, 
respectively. The discrete points show growth rates calculated by direct linear numerical 
simulation using the gyrokinetic code {\tt AstroGK} \cite{Numata10}. 
{\em Right panel:} Contour plot of $\gamma/(qS/\ee)$ vs.\ $\kpar\vth/(qS/\ee)$ 
and $k_y\rhoi$. Only positive values are plotted, black means $\gamma<0$.
The red curve shows the stability boundary \eqref{eq:marg}. 
The white line shows the boundary \eqref{eq:soundwedge} 
between PVG modes (above it) and sound waves (below it).
}
\end{figure}

Some further (elementary) analytical considerations of the PVG 
instability are relegated to \apref{app:pvg}. Here it will suffice 
to notice that if the (dimensional) growth rate 
$\gamma=\Im\,\omega$ is scaled by $qS/\ee$ and $\kpar$ 
by $(qS/\ee)/\vth$, their mutual dependence is universal for all values of the 
velocity shear or $q/\ee$, namely, 
\beq
\gamma = \frac{q S}{\ee}\,f\lt(k_y\rhoi,\frac{\kpar\vth}{qS/\ee}\rt).
\label{eq:gamma_gen_pvg}
\eeq 
Once this rescaling is done, the dispersion relation \eqref{eq:pvg} no longer 
contains any parameters (except $\tau/Z$, which we can safely take to be order unity). 
The growth rate, obtained via numerical solution 
of \eqref{eq:pvg} with $\tau/Z=1$, is plotted in \figref{fig:pvg}. 
The maximum growth rate is $\gamma_\mathrm{max}\approx 0.10(qS/\ee)$. 
This peak value is reached when $k_y\rhoi\approx1.0$ and $\kpar\approx 0.10(qS/\ee)/\vth$, 
i.e., at $(q/\ee)\zS\approx 10$. 

The conclusion is that, at least in the initial stage 
of their evolution, plasma fluctuations in a significant part of the 
wavenumber space ($k_y$ and $\kpar$ are constrained by \eqref{eq:marg}) 
are amplified by the PVG. 
This amplification does not, however, go on for a long time as the approximation 
we adopted to derive the dispersion relation \eqref{eq:pvg} breaks down when 
$\zS t\sim k_y\rhoi$ (or $St\sim 1$ if the dimensional units of time are restored)
--- this gives $\gamma t\sim 0.1 (q/\ee)$, so, 
realistically, after barely one exponentiation. 
The key question is what happens after that. 
We will see shortly that all modes will eventually decay and 
that the fastest initially growing modes are in fact not quite 
the ones that will grow the longest or get maximally amplified. 

\subsection{Long-time limit: transient growth}
\label{sec:longt}

Let us now investigate the long-time limit, $\zS t\gg 1$, $k_y\rhoi$ 
(or, in dimensional form, $St\gg1$, $k_y\rhoi St\gg1$).  
In this limit, the kernel involving the Bessel function in \eqref{eq:inteq_pvg} 
simplifies considerably: we have $\lambda\approx\zS^2t^2/2\gg1$, 
$\lambda'\approx\zS^2(t-\tbar)^2/2\gg1$ and so 
\beq
\label{eq:Gamma0_longt}
\Gamma_0(\lambda,\lambda')\approx 
\frac{e^{-(\sqrt{\lambda}-\sqrt{\lambda'})^2/2}}{\sqrt{2\pi\sqrt{\lambda\lambda'}}}
\approx \frac{e^{-\zS^2\tbar^2/4}}{\sqrt{\pi}|\zS|t}.
\eeq
Working to the lowest nontrivial order in $1/t$, we can now rewrite 
\eqref{eq:inteq_pvg} as follows
\beq
\fl
\lt(1+\frac{\tau}{Z}\rt)|\zS|t\tphi(t) 
= \frac{1}{2\sqrt{\pi}}\int_0^\infty d\tbar\,\tbar\,e^{-(\zS^2+1)\tbar^2/4}
\lt(\frac{q}{\ee}\,\zS-1\rt)\tphi(t-\tbar).
\label{eq:longt}
\eeq
We will seek a solution to this equation in the form 
\beq
\tphi(t) = \tphi_0 \exp\lt[\int_0^t dt'\gbar(t')\rt],
\label{eq:phit}
\eeq
where $\gbar(t)=\gamma(t)/|\kpar|\vth$ is the effective time-dependent 
growth rate (nondimensionalised) and $\gamma(t)$ the dimensional version 
of it (remember that time is scaled by $|\kpar|\vth$). 
Because of the exponential in the kernel, the memory of the time-history 
integral in the right-hand side of \eqref{eq:longt} is limited, so 
$\tbar\ll t$ and we will be able to make progress by expanding
\beq
\tphi(t-\tbar) = \tphi_0 \exp\lt[\int_0^t dt'\gbar(t') - \tbar\gbar(t) 
+ \frac{\tbar^2}{2}\,\gbar'(t) + \dots\rt].
\label{eq:phi_exp}
\eeq
We will assume that this expansion can be truncated; the resulting 
solution will indeed turn out to satisfy 
$\gbar'(t)\ll1$, with all higher-order terms even smaller. 
Substituting \eqref{eq:phi_exp} into \eqref{eq:longt}, we obtain an implicit 
transcendental equation for $\gbar(t)$: 
\beq
\lt(1+\frac{\tau}{Z}\rt)|\zS|t 
= \frac{1}{2\sqrt{\pi}}\int_0^\infty d\tbar\,\tbar\,e^{-\tbar\gbar(t)-(\zS^2+1)\tbar^2/4}
\lt(\frac{q}{\ee}\,\zS-1\rt).
\label{eq:zeta}
\eeq
This equation can be written in a compact form by invoking once again  
the plasma dispersion function \eqref{eq:Z}: denoting 
$\tg(t)=\gbar(t)/\sqrt{\zS^2+1}$, we get
\beq
\lt(1+\frac{\tau}{Z}\rt)\sqrt{\pi}(\zS^2+1)|\zS|t =
\lt(\frac{q}{\ee}\,\zS-1\rt) [1+i\tg(t)\Zfn(i\tg(t))] 
\label{eq:tz}
\eeq
--- effectively, a time-dependent dispersion relation, reminiscent 
of the PVG dispersion relation \eqref{eq:pvg}. 

\subsubsection{Transient growth.}
\label{sec:trans}

First of all, 
it is immediately clear from \eqref{eq:zeta} that as time increases, 
(the real part of) the effective time-dependent growth rate $\gbar(t)$  
must decrease and indeed go negative because 
the right-hand side has to keep up with the increasing left-hand side.
Therefore, fluctuations will eventually decay. However, 
if $\Re\,\gbar(t)$ is positive for some significant initial period of time, 
there can be a substantial transient amplification. 

We can determine 
the time $t_0$ when the transient growth ends by setting 
$\Re\,\tg(t_0) = 0$ in \eqref{eq:tz}. We immediately find that 
$\Im\,\tg(t_0) = 0$ as well and that 
\beq
t_0 =\frac{(q/\ee)\zS-1}{(1+\tau/Z)\sqrt{\pi}(\zS^2+1)|\zS|}.
\label{eq:t0}
\eeq
The dependence of $t_0$ on $k_y$ and $\kpar$ 
--- via $\zS=Sk_y\rhoi/\kpar\vth$ 
and via the time normalisation factor of $|\kpar\vth|$ 
--- tells us which modes grow longest. The interesting question, however, 
is rather which modes get maximally amplified 
during this transient growth. 

\begin{figure}[t]
\centerline{\epsfig{file=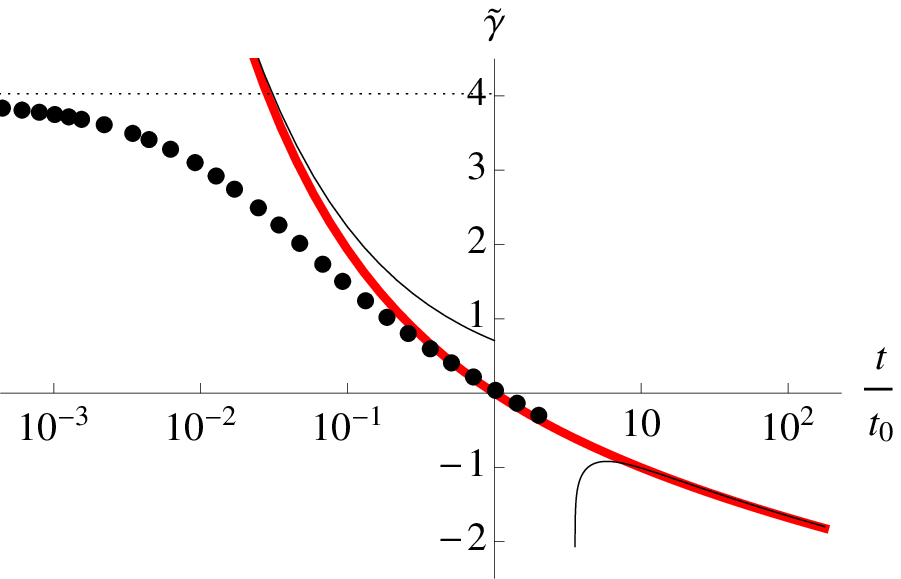,width=3.5in}}
\caption{\label{fig:gamma} 
Time evolution of the effective growth rate: the red (bold) line is   
$\tg(t) = \gamma(t)/|\kpar|\vth\sqrt{\zS^2+1}$ 
obtained as a numerical solution of \eqref{eq:pvgt} and 
plotted vs.\ $t/t_0$ (the time axis is logarithmic in base 10). 
The black (thin) lines are the asymptotics 
\eqref{eq:growth} and \eqref{eq:decay_precise}.
The discrete points show the time evolution of the effective growth rate 
obtained in a direct linear numerical simulation using the gyrokinetic 
code {\tt AstroGK} \cite{Numata10} with
$1/\Ln=1/\LT=0$, $q/\ee=50$, $\tau/Z=1$, $k_y\rhoi=1$, $\kpar\vth/S=0.5$. 
The short-time-limit PVG growth rate 
for this case (obtained by solving \eqref{eq:pvg}) is shown as a dotted 
horizontal line. The time is normalised using \eqref{eq:t0} for $t_0$.
The time evolution of the perturbation amplitude for this case is 
shown in \figref{fig:trgrowth}.} 
\end{figure}

\subsubsection{Maximal amplification.}
\label{sec:Nmax}

The total amplification factor 
is given by $e^N$, where $N = \int_0^{t_0}dt\gbar(t)$ is the number 
of exponentiations experienced by the mode during its growth period. 
In order to determine this, we need to know the time evolution 
of $\gbar(t)$ up to $t=t_0$. 
Using \eqref{eq:t0}, it is convenient to rewrite \eqref{eq:tz} as follows 
\beq
\frac{t}{t_0} = 1 + i\tg\,\Zfn(i\tg).
\label{eq:pvgt}
\eeq 
When $t\ll t_0$, the solution is found by expanding the 
plasma dispersion function in $\tg\gg1$: 
\beq
\frac{t}{t_0}\approx \frac{1}{2\tg^2}
\quad\Rightarrow\quad
\tg\approx \sqrt{\frac{t_0}{2t}}.
\label{eq:growth}
\eeq
This asymptotic is not valid when $t$ approaches $t_0$. 
More generally, \eqref{eq:pvgt} has a solution 
$\tg=\tg(t/t_0)$, whose functional form is independent of 
any parameters of the problem. It is plotted in \figref{fig:gamma} 
together with the asymptotic \eqref{eq:growth} and with a direct 
numerical solution showing how the transition between the 
short-time (\secref{sec:pvg}) and long-time limits occurs. 
The amplification exponent is easily found:
\beq
\fl
N = \int_0^{t_0}dt\gbar(t) 
= t_0\sqrt{\zS^2+1}\int_0^1 d\xi\,\tg(\xi)
\approx 0.45\,\frac{(q/\ee)\zS-1}{(1+\tau/Z)\sqrt{\zS^2+1}\,|\zS|},
\label{eq:N}
\eeq
where we have used \eqref{eq:t0} for $t_0$ and computed the 
integral numerically. It is clear from \eqref{eq:growth} that the 
integral converges on its lower limit and is not dominated by it, 
so it does not matter that we cannot technically use \eqref{eq:pvgt} 
for short times. 

\begin{figure}[t]
\centerline{\epsfig{file=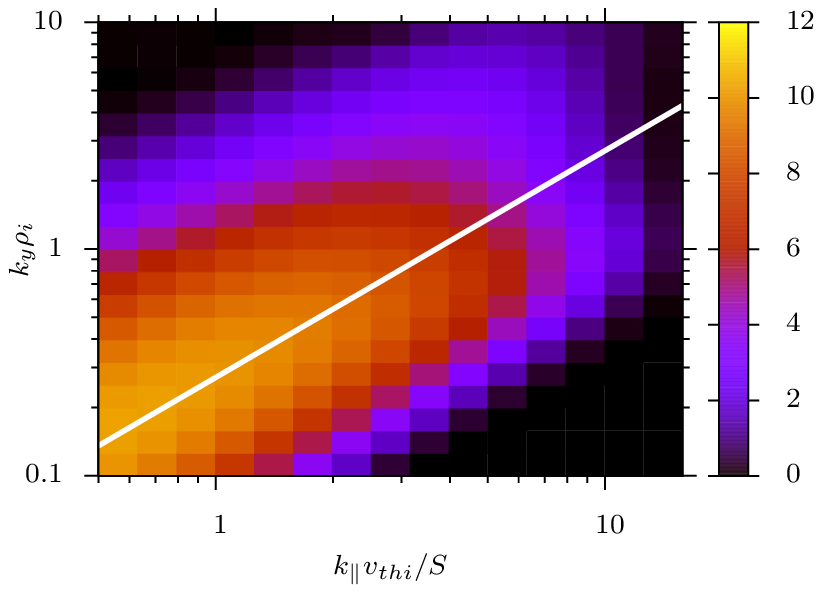,width=2.75in}
\hskip0.25in
\epsfig{file=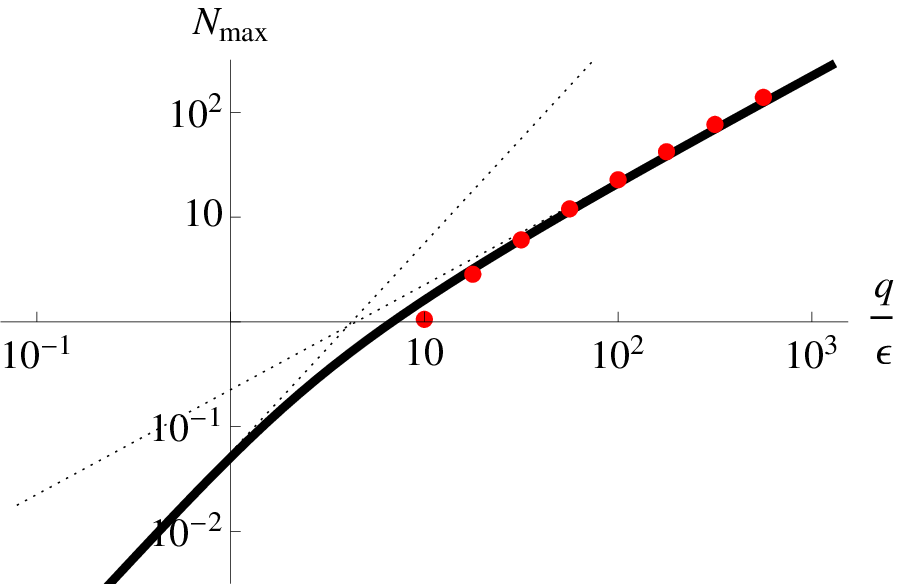,width=3in}}
\caption{\label{fig:kspace} 
{\em Left panel:} The amplification exponent $N$ vs.\ $\kperp$ and $\kpar$ 
in the same {\tt AstroGK} simulation as used to produce the 
discrete points in \figref{fig:gamma}. The straight line 
shows the relationship between the wavenumbers given by 
the second formula in \eqref{eq:Nmax}.
{\em Right panel:} The maximum amplification exponent $\Nmax$ 
[$N$ given by \eqref{eq:N}, maximised with respect to $\zS$] vs.\ $q/\ee$.
The dotted lines show the $q/\ee\gg1$ asymptotic [see \eqref{eq:Nmax}] 
and the $q/\ee\ll1$ asymptotic, the latter straightforwardly 
obtained from \eqref{eq:N}: $\zS\approx2(\ee/q)$, $\Nmax\approx0.11(q/\ee)^2/(1+\tau/Z)$ 
(but this is purely formal because the long-time conditions $St_0\gg1$ 
and $k_y\rhoi St_0\gg1 $ will be broken in this regime, except for extremely long wavelengths).
The discrete points show $\Nmax$ obtained via an {\tt AstroGK} numerical parameter 
scan: $\kpar\vth/S=0.5$, varying $\kperp\rhoi$ and holding all
other parameters fixed as in \figref{fig:gamma} 
(note that the asymptotic results do not in fact depend on $\kpar$ or $S$, although 
the quality of the long-time asymptotic does).} 
\end{figure}

According to \eqref{eq:N}, $N$ depends on both wavenumbers via $\zS$ only 
(a plot of this dependence will be given in \figref{fig:N}). 
Assuming $q/\ee\gg1$, we find that the amplification exponent is maximised 
for $\zS \approx (\ee/q)^{1/3}$, giving\footnote{Note that this is well outside 
of the wavenumber domain populated by the damped sound waves, 
$\zS<\ee/q$ (see \apref{app:pvg}).} 
\beq
\Nmax \approx 
0.45\,\frac{q/\ee}{1+\tau/Z}
\quad {\rm for}\quad 
k_y\rhoi\approx \lt(\frac{\ee}{q}\rt)^{1/3}\frac{\kpar\vth}{S}.
\label{eq:Nmax}
\eeq
These results are illustrated in the left panel of \figref{fig:kspace}.
The amplification time for the maximally amplified modes 
identified in \eqref{eq:Nmax} is, from \eqref{eq:t0}, 
\beq
t_0 \approx \frac{q/\ee}{|\kpar|\vth(1+\tau/Z)\sqrt{\pi}}
\approx \frac{(q/\ee)^{2/3}}{|Sk_y\rhoi|(1+\tau/Z)\sqrt{\pi}},
\label{eq:t0_max} 
\eeq
where we have restored dimensions to make explicit the dependence 
of $t_0$ on $\kpar$.

Thus, we have learned that an entire family of modes, 
characterised by a particular (linear) relationship between $k_y$ and $\kpar$,
given in \eqref{eq:Nmax}, will eventually enjoy the same 
net amplification, even though, as follows from the results 
of \secref{sec:pvg}, they were not the fastest initially 
growing modes and some within this equally amplified family 
started off growing more slowly than others or even decaying. 
The more slowly growing modes are the longer-wavelength ones and,  
according to \eqref{eq:t0_max}, they compensate for their 
sluggishness with longer growth times. 

Note that \eqref{eq:t0_max} confirms that the long-time limit is 
analytically reasonable because for large~$q/\ee$, 
\eqref{eq:t0_max} formally satisfies $k_y\rhoi St_0\sim(q/\ee)^{2/3}\gg1$ 
and also $St_0\gg1$, provided $\kpar\vth/S\ll q/\ee$. 
The latter condition is marginally broken by the 
fastest initially growing modes: indeed,  
in \secref{sec:pvg}, we saw that they had $\kpar\vth/S\sim 0.1 q/\ee$, 
so for them, $St_0\sim 3$, not really a large number.
For all longer-wavelength modes, $t_0$ is safely 
within the domain of validity of the long-time limit.

Note also that, as we show at the end of \apref{app:pvg}, the time-dependent 
dispersion relation \eqref{eq:pvgt} and its consequences derived above can be 
obtained directly from the PVG-instability dispersion relation \eqref{eq:pvg} 
simply by restoring its $k_x$ dependence, setting $k_x=Sk_yt$, taking the 
short-wavelength and long-time limit ($k_x\rhoi\gg1$, $St\gg1$), 
and assuming $\ee/q\ll\zS\ll1$. This calculation underscores the fundamental
simplicity of the physics of the transient amplification: perturbations 
initially destabilised by the PVG are eventually swept by the perpendicular 
velocity shear into a stable region of the wavenumber space. 

\subsubsection{Limits on short and long wavelengths.}
\label{sec:UV}

We have seen that modes with parallel wavenumbers up to 
$\kpar\vth/S\sim q/\ee$ can be transiently amplified. 
From \eqref{eq:Nmax}, we conclude that of these, 
the maximally amplified ones will have perpendicular wavenumbers 
up to $k_y\rhoi\lesssim (q/\ee)^{2/3}$, i.e., 
$k_y\rhoi$ can be relatively large --- unlike in the short-time limit 
treated in \secref{sec:pvg}, where the modes with 
$k_y\rhoi\sim1$ grew the fastest (although large $k_y\rhoi$ were also 
unstable).

It should be understood that, while there is no ultraviolet 
cutoff in our theory that would limit the wavenumbers of the growing modes 
(in either direction), 
such a cutoff does of course exist in any real system. 
In the parallel direction, those $\kpar$ that were strongly damped 
in the short-time limit (see \eqref{eq:marg}) are unlikely to recover 
in the long-time limit. In the perpendicular direction, the cutoff 
in $k_y\rhoi$ will come from the collisional damping, which, in 
gyrokinetics, contains a spatial diffusion (see, e.g., \cite{Abel08}), 
and from the electron Landau damping, which we have lost by using 
the Boltzmann electron response (see \eqref{eq:quasineut_sh}) and 
which should wipe out large $k_y$ and $\kpar$. 

On the infrared (long-wavelength) side of the spectrum, we have no 
cutoffs either. In a slab, these would be provided by the dimensions 
of the periodic box. In a real plasma, the cutoffs are set by 
the scales at which the system can no longer be considered 
homogeneous (in a tokamak, these are the equilibrium-gradient scale lengths 
and the minor radius for the perpendicular scales and the connection 
length $qR$ for the parallel scales; we will need these considerations 
to fix transport scaling in \secref{sec:turb}). 

\subsubsection{Significant amplification threshold.}
\label{sec:threshold}
If we maximise \eqref{eq:N} without assuming $q/\ee\gg1$ (with the caveat 
that the long-time limit asymptotics are at best marginally valid then), 
we obtain a more general curve than \eqref{eq:Nmax}, plotted in the right 
panel of \figref{fig:kspace}. We may define a critical threshold for 
significant amplification: $\Nmax=1$ when $q/\ee\approx7$. 
The role of this threshold will be discussed in \secref{sec:marginal}. 

\subsubsection{Long-time decay.}
\label{sec:decay}

Finally, we obtain the long-time asymptotic decay law. 
Let us seek a solution of \eqref{eq:pvgt} 
such that $t\gg t_0$ and $\tg\ll-1$. Then 
\beq
\frac{t}{t_0} \approx 2\sqrt{\pi}\,|\tg| e^{\tg^2}
\quad\Rightarrow\quad
\tg\approx -\sqrt{\ln\frac{t}{t_0}}.
\label{eq:decay}
\eeq
Since $\tg$ is only root-logarithmically large, the quality of this 
asymptotic is rather poor. If we insist on a more precise decay law, 
we can retain small corrections in \eqref{eq:decay} and 
get what turns out, upon a numerical test,
to be a reasonably good approximation (see \figref{fig:gamma}):
\beq
\tg \approx -\sqrt{\ln\frac{t/(2\sqrt{\pi}t_0)}{\sqrt{\ln(t/2\sqrt{\pi}t_0)}}}.
\label{eq:decay_precise}
\eeq
For the maximally amplified modes (see \eqref{eq:Nmax}), 
the dimensional damping rate is $\gamma(t)\approx|\kpar|\vth\tg(t)$ 
with $t_0$ given by \eqref{eq:t0_max}.  
This tells us is that the decay 
is just slightly faster than exponential at the rate of order~$S$. 
The longest-wavelength modes decay the slowest, after having 
being amplified the longest.  

\section{Solution including ITG}
\label{sec:incl_itg}

Let us now generalise the results obtained in \secref{sec:pure_pvg} to 
include non-zero (i.e., non-negligible) density and temperature gradients. 
This means that we restore the terms involving $\zn$ and $\eta_i$ in 
the general integral equation \eqref{eq:inteq}. 

\subsection{Short-time limit: the ITG-PVG dispersion relation}
\label{sec:itg}

The short-time limit, introduced at the beginning of \secref{sec:pvg}
for the case of pure PVG, is treated in an analogous fashion for the 
general ITG-PVG case.
An analysis of the solutions of the resulting dispersion relation is 
useful in that its results assist 
physical intuition in ways relevant for some of the forthcoming 
discussion, but it is not strictly necessary for us to have them 
in order to work out how transient growth happens in the 
presence of the ITG drive. We have therefore relegated 
this analysis to \apref{app:itg}. 

\subsection{Long-time limit}

We now continue in the same vein as in \secref{sec:longt} and 
consider the long-time limit ($St\gg1$, $k_y\rhoi St\gg1$), in which 
we can simplify the kernels involving the Bessel functions in \eqref{eq:inteq} 
by using \eqref{eq:Gamma0_longt} 
and also 
\beq
\Lambda(\lambda,\lambda')\approx \frac{1}{2} - \frac{(\sqrt{\lambda}-\sqrt{\lambda'})^2}{2}
\approx \frac{1}{2} - \frac{\zS^2\tbar^2}{4}. 
\eeq
This allows us to rewrite \eqref{eq:inteq} in the form that 
generalises \eqref{eq:longt}: 
\bea
\nonumber
\fl
\lt(1+\frac{\tau}{Z}\rt)|\zS|t\tphi(t) 
&=& \frac{1}{\sqrt{\pi}}\int_0^\infty d\tbar\,e^{-(\zS^2+1)\tbar^2/4}
\lt\{\lt(\frac{q}{\ee}\,\zS-1\rt)\frac{\tbar}{2}\rt.\\ 
&&\qquad\qquad-\lt. i\zn\lt[1-\eta_i\lt(\frac{1}{2} + \frac{(\zS^2+1)\tbar^2}{4}\rt)\rt]\rt\}
\tphi(t-\tbar).
\label{eq:longt_itg}
\eea
As in \secref{sec:longt}, we 
seek a solutions to this equation in the form \eqref{eq:phit}, 
taking $\tbar\ll t$ and expanding the delayed potential under 
the integral according to \eqref{eq:phi_exp}. 
The result is the generalised form of \eqref{eq:zeta}:
\bea
\nonumber
\fl
\lt(1+\frac{\tau}{Z}\rt)|\zS|t 
&=& \frac{1}{\sqrt{\pi}}\int_0^\infty d\tbar\,e^{-\tbar\gbar(t)-(\zS^2+1)\tbar^2/4}
\lt\{\lt(\frac{q}{\ee}\,\zS-1\rt)\frac{\tbar}{2}\rt.\\ 
&&\qquad\qquad-\lt. i\zn\lt[1-\eta_i\lt(\frac{1}{2} + \frac{(\zS^2+1)\tbar^2}{4}\rt)\rt]\rt\}.
\label{eq:zeta_itg}
\eea
Using again the plasma dispersion function \eqref{eq:Z} to 
express the time integrals and introducing the complex scaled frequency 
$\tz(t)=i\gbar(t)/\sqrt{\zS^2+1}$, we get
\bea
\fl
\nonumber
\lt(1+\frac{\tau}{Z}\rt)\sqrt{\pi(\zS^2+1)}|\zS|t &=&  
\lt[\frac{(q/\ee)\zS-1}{\sqrt{\zS^2+1}} - \eta_i\zn\tz(t)\rt]
[1+\tz(t) \Zfn(\tz(t))]\\ 
&&+\,\, (\eta_i-1)\,\zn \Zfn(\tz(t)),
\label{eq:tz_itg}
\eea
a time-dependent dispersion relation, which is the generalisation of \eqref{eq:tz}. 

\subsubsection{Transient growth.} The general argument that
the real part of $\gbar(t)$ (i.e., the effective time-dependent growth rate) 
must eventually decrease and so fluctuations will, in the end, decay, applies 
to \eqref{eq:zeta_itg} similarly to the way it did to \eqref{eq:zeta} 
(see \secref{sec:trans}), although this decay need not (and, as we will see, 
will not) be monotonic. 
The time $t_0$ when the transient growth ends is now determined as follows.  
Let $\Im\,\tz(t_0) = 0$ and $\tz(t_0)=\tz_0$ (real!). Then from  
\eqref{eq:tz_itg} taken at $t=t_0$, we find the real frequency $\tz_0$ by demanding 
that the imaginary part of the right-hand side vanish --- this means that 
the coefficient in front of $\Zfn(\tz_0)$ must be zero, because, for real $\tz_0$, 
$\Im\,\Zfn(\tz_0) = \sqrt{\pi}\, e^{-\tz_0^2}$. This condition gives
\beq
\eta_i\zn\tz_0^2 - \frac{(q/\ee)\zS-1}{\sqrt{\zS^2+1}}\,\tz_0 - (\eta_i-1)\zn = 0,
\eeq
whence
\beq
\tz_0 = \frac{(q/\ee)\zS-1 \pm \sqrt{[(q/\ee)\zS-1]^2 + 4\eta_i(\eta_i-1)\,\zn^2\,(\zS^2+1)}}{2\eta_i\zn\sqrt{\zS^2+1}}.
\label{eq:freq}
\eeq
Substituting this solution into the real part of \eqref{eq:tz_itg} and taking 
advantage of the already enforced vanishing of the coefficient in front 
of $\Zfn(\tz_0)$, we get 
\beq
t_0 = \frac{(q/\ee)\zS - 1 + \sqrt{[(q/\ee)\zS - 1]^2 
+ \eS^2\,(1-1/\eta_i)\,\zS^2\,(\zS^2+1)}}{2(1+\tau/Z)\sqrt{\pi}(\zS^2+1)|\zS|},
\label{eq:t0_itg}
\eeq
where we have replaced $\eta_i^2\zn^2 = \eS^2\zS^2/4$ 
with $\eS = \vth/\LT S$ a new parameter that measures 
the strength of the ITG drive relative to the velocity shear.
Note that we picked the ``$-$'' mode in \eqref{eq:freq} because the ``+'' mode 
is not amplified ($t_0<0$, assuming $\eta_i>1$). 
\Eqref{eq:t0_itg} is the generalisation of \eqref{eq:t0}, 
to which it manifestly reduces when $\eS=0$ and with which it shares the property 
that the transient growth time depends on $k_y$ and $\kpar$ 
only via $\zS$ and the time normalisation factor $|\kpar|\vth$. 

\subsubsection{General dispersion relation.} We can now recast 
the general ITG-PVG case in a form that shows explicitly how it reduces 
to the case of pure PVG drive studied in \secref{sec:longt}. First we note 
that the transient growth termination time \eqref{eq:t0_itg} can be 
rewritten as
\beq
\fl
t_0 = \frac{1}{2}\lt[1 + \sigma
\sqrt{1+ \lt(1-\frac{1}{\eta_i}\rt)\chi^2}\rt]\tpvg,
\qquad
\chi = \frac{\eS\,\zS \sqrt{\zS^2+1}}{(q/\ee)\zS - 1},
\label{eq:chi}
\eeq
where $\tpvg$ is given by \eqref{eq:t0}, $\eS\zS=2\eta_i\zn$, $\eS=\vth/\LT S$
and $\sigma = {\rm sgn}[(q/\ee)\zS - 1]$.\footnote{If $(q/\ee)\zS < 1$, 
$\tpvg<0$ and $\chi<0$, but the ITG mode can still have transient growth, $t_0>0$.}  
Then the time-dependent dispersion relation \eqref{eq:tz_itg} can be manipulated into 
the following form:
\beq
\fl
\lt[1 + \sigma\sqrt{1+ \lt(1-\frac{1}{\eta_i}\rt)\chi^2}\rt]
\frac{t}{t_0} =
(2 - \chi\,\tz) [1+\tz\,\Zfn(\tz)] 
+ \lt(1-\frac{1}{\eta_i}\rt)\chi\,\Zfn(\tz).
\label{eq:itgt}
\eeq
The analogous equation for the case of pure PVG drive, 
\eqref{eq:pvgt}, is recovered when $\chi\ll1$, which 
means $\eS\ll q/\ee$ (cf.\ \eqref{eq:pure_pvg}) and $\eS\zS\ll1$. 
In this limit, the behaviour of fluctuations in the presence 
of both PVG and ITG drives is well described by 
the results of \secref{sec:longt}. Before discussing the general case, 
it is useful to consider the opposite extreme of weak velocity shear.

\subsection{Case of weak shear} 
\label{sec:weakS}

\begin{figure}[t]
\centerline{\epsfig{file=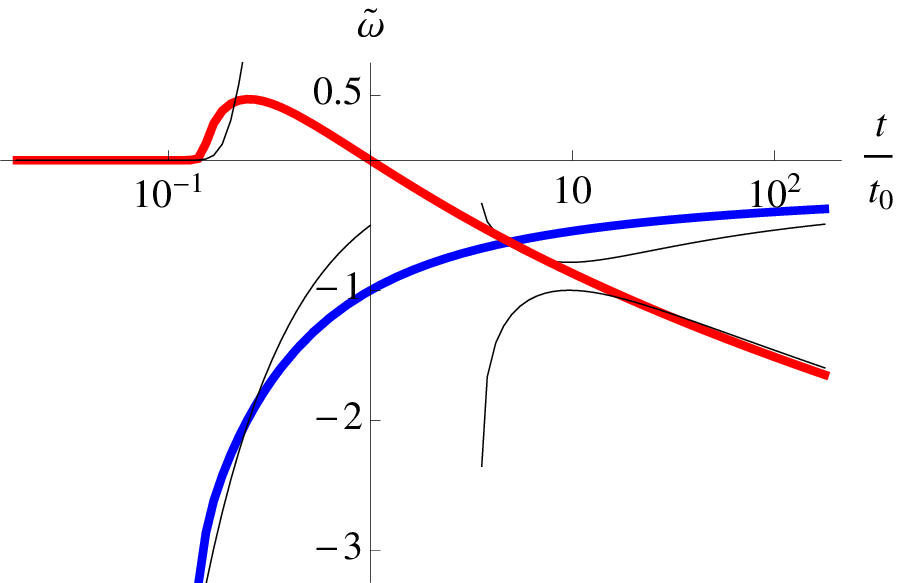,width=3.5in}}
\caption{\label{fig:weakS} 
Time evolution of the effective growth rate and frequency: 
the red (upper bold) line 
is  $\tg(t) = \Im\,\omega(t)/|\kpar|\vth\sqrt{\zS^2+1}$ 
and the blue (lower bold) line is 
$\Re\,\omega(t)/|\kpar|\vth\sqrt{\zS^2+1}$, both   
obtained as a numerical solution of \eqref{eq:weakS} and 
plotted vs.\ $t/t_0$ (the time axis is logarithmic$_{10}$). 
The black (thin) lines show the growth rate and frequency 
given by the asymptotics \eqref{eq:growth_weakS} 
and \eqref{eq:decay_itg_precise} (the latter taken in the limit 
$\chi\to\infty$ and $\eta_i\to\infty$).} 
\end{figure}

Let $\eS\gg q/\ee$ and $\eS\zS\gg1$, 
so $\chi\gg1$ (note that the same limit is also achieved for $\zS\to\infty$). 
Then the $\chi$ dependence falls out of \eqref{eq:itgt}: 
\beq
\frac{t}{t_0} = - \tz [1+\tz\,\Zfn(\tz)] + \Zfn(\tz),
\label{eq:weakS}
\eeq
where we have discarded the $1/\eta_i$ terms by assuming $\eta_i\gg1$. 
The transient growth termination time in this limit is, from \eqref{eq:chi}, 
\beq
t_0 = \frac{1}{2}\,\chi\,\tpvg = \frac{\eS}{2(1+\tau/Z)\sqrt{\pi(\zS^2+1)}}.
\label{eq:t0_weakS}
\eeq
When $t\ll t_0$, we find the solution of \eqref{eq:weakS} by expanding in 
$\tz\gg1$. It turns out that it consists of a large real frequency and 
an exponentially small growth rate: \eqref{eq:weakS} becomes
\beq
\fl
\frac{t}{t_0} \approx - \frac{1}{2\tz} - i\sqrt{\pi}\,\tz^2\,e^{-\tz^2}
\quad\Rightarrow\quad
\tz \approx -\frac{t_0}{2t} + 
i\,2\sqrt{\pi}\lt(\frac{t_0}{2t}\rt)^4 e^{-(t_0/2t)^2}.
\label{eq:growth_weakS}
\eeq
More generally, for finite values of $t/t_0$, 
the solution of \eqref{eq:weakS} is $\tz=\tz(t/t_0)$, with a functional 
form independent of the parameters of the problem. This solution it 
plotted in \figref{fig:weakS}. It turns out that at 
$t\approx 0.15\, t_0$, the growth rate increases sharply, 
reaches a finite maximum and then decreases towards zero, 
which it reaches at $t=t_0$, whereupon growth turns to decay.\footnote{This 
implies that perturbations first grow due to the ITG-PVG 
instability (at $St\ll1$), then slow down to exponentially small 
growth rates, then (at $St\gg1$) grow vigorously again before finally starting 
to decay at $t=t_0$. The intermediate period of virtually zero growth, 
which is a feature both of the weak-shear regime and of the general 
case (see \secref{sec:finiteS}) and 
may appear strange at first glance, can be traced to the 
$k_x$ dependence of ITG-PVG growth rates at long parallel wavelengths
--- this is explained in \apref{app:itg}.}

Thus, there is a period of strong transient amplification, which lasts 
for a finite fraction of time $t_0$. The amplification exponent is 
\beq
N = \int_0^{t_0}dt\gbar(t) 
= t_0\sqrt{\zS^2+1}\int_0^1 d\xi\,\Im\,\tz(\xi)
\approx 0.057\,\frac{\eS}{1+\tau/Z},
\label{eq:N_weakS}
\eeq
where we have used \eqref{eq:t0_weakS} and calculated the value of the 
integral under the curve in \figref{fig:weakS} numerically
(note that since the growth rate is exponentially small at $t\ll t_0$, 
the precise lower integration limit is irrelevant). 
Remarkably, unlike in the case of the PVG drive (see \eqref{eq:Nmax}), 
the amplification exponent has no wavenumber dependence 
at all. Also unlike in the PVG case, it does depend on the shear and on the 
temperature gradient: $N\propto\eS=\vth/\LT S$. 

To recapitulate, 
we have found that, at low velocity shear, all modes are amplified by a large (and the same) 
factor before decaying eventually. Their transient amplification time
is given by \eqref{eq:t0_weakS}. Restoring dimensions, \eqref{eq:t0_weakS} 
and \eqref{eq:N_weakS} are
\beq
\fl
t_0 \approx \frac{\vth/(\LT S)}{2(1+\tau/Z)\sqrt{\pi(S^2k_y^2\rhoi^2 + \kpar^2\vth^2)}},
\quad
N \approx 0.057\,\frac{\vthi/(\LT S)}{1+\tau/Z}.
\label{eq:t0_weakS_dim}
\eeq
The transient growth lasts for a very long time at low $S$ and the 
longest-growing modes are the long-wavelength ones. 
The limit $S\to 0$ is singular in the sense that for arbitrarily small but 
non-zero $S$ all modes eventually decay, while for $S=0$, the indefinitely 
growing linear ITG instability is recovered ($t_0=\infty$, $N=\infty$). 

We have already made the point (in \secref{sec:UV}) that while our theory 
does not limit the transiently growing wavenumbers from above, 
a fuller description of the plasma will. 

\begin{figure}[t]
\centerline{\epsfig{file=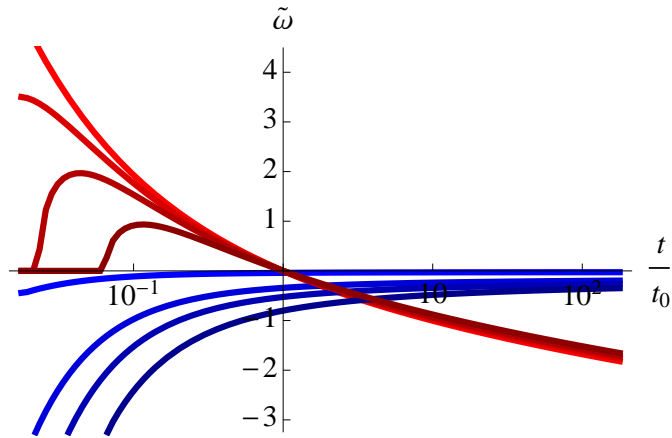,width=3.5in}}
\caption{\label{fig:gamma_itg} Effective normalised growth rates $\Im\,\tz(t/t_0)$ (red, top) 
and frequencies $\Re\,\tz(t/t_0)$ (blue, bottom) 
obtained via numerical solution of \eqref{eq:itgt} with 
$\eta_i=5$ and $\chi=0.1,1,2,10$ (from top/lighter to bottom/darker curves).
See \figref{fig:gamma_chi1} for a more detailed depiction of the 
$\chi=1$ case (in \apref{app:longt}, where the functional 
form of these curves is derived analytically).} 
\end{figure}

\begin{figure}[t]
\centerline{\parbox{3in}{\epsfig{file=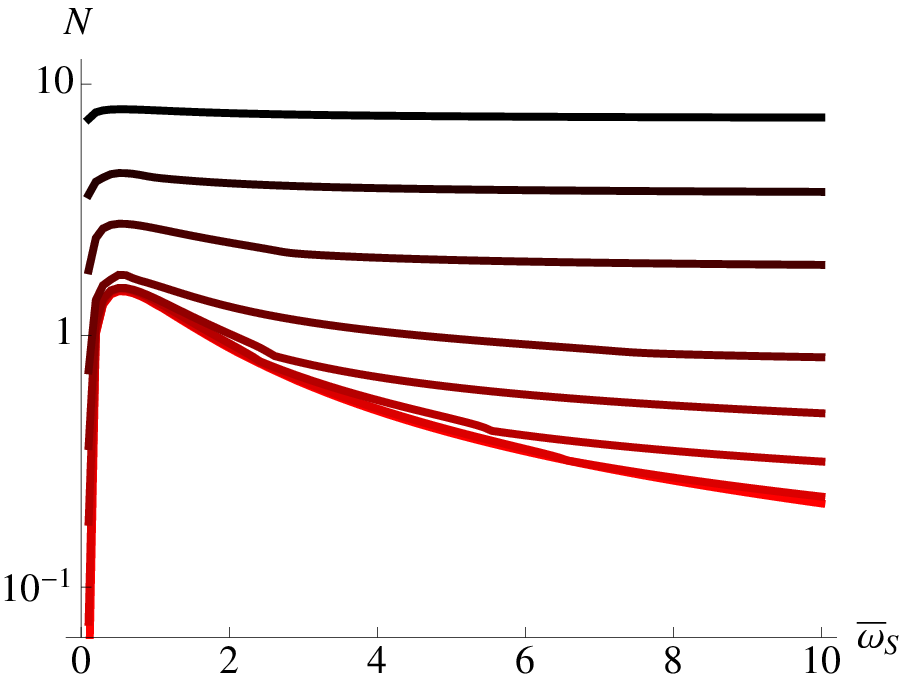,height=2.2in}}
\parbox{3in}{\epsfig{file=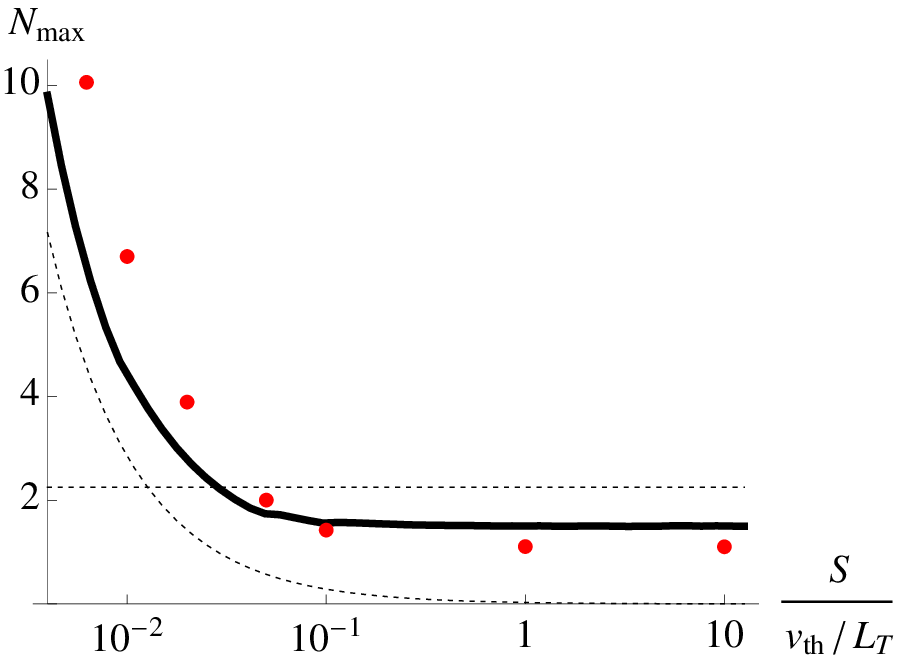,height=2.2in}}}
\caption{\label{fig:N} The amplification exponent $N$, given by \eqref{eq:N_itg} with 
$\tz$ the solution of \eqref{eq:itgt} for $\eta_i=5$, $q/\ee=10$ and $\tau/Z=1$.  
{\em Left panel:} $N$ vs.\ $\zS$ for 
$\eS=1,2,5,10,20,50,100,200$ (from bottom to top curves, in darkening shades of red). 
The numerical results for the same cases are shown in \figref{fig:N2D}. 
The $\eS=0$ case (pure PVG drive) was virtually indistinguishable 
from $\eS=1$ when plotted (not shown here). 
The lowest-$\eS$ behaviour is well described by \eqref{eq:N}, 
the highest-$\eS$ by \eqref{eq:N_weakS} (in the latter case, except 
for corrections associated with finite $\eta_i$, which are easy 
to compute if they are required).
{\em Right panel:} The maximal amplification exponent $N$ vs.\ 
the normalised shear ($\eS^{-1}=S\LT/\vth$). 
The maximum $N(\zS)$ is reached at $\zS\approx 0.54$, independently of $\eS$. 
The dotted lines are the asymptotics 
\eqref{eq:Nmax} and \eqref{eq:N_weakS}. The finite offsets between the 
asymptotics and the exact curve are due to the fact that the asymptotics 
were calculated in the limits $\eta_i\gg1$ and $q/\ee\gg1$, while for the 
exact solution we used relatively moderate values of these parameters.
The discrete points show $\Nmax$ obtained via an {\tt AstroGK} 
numerical parameter scan varying $\kperp\rhoi$ and $S$ (i.e., $\eS$) while 
holding $\kpar\LT=0.02$ and the rest of the parameters fixed at the same values 
as quoted above.} 
\end{figure}

\begin{figure}[p!]
\centerline{\epsfig{file=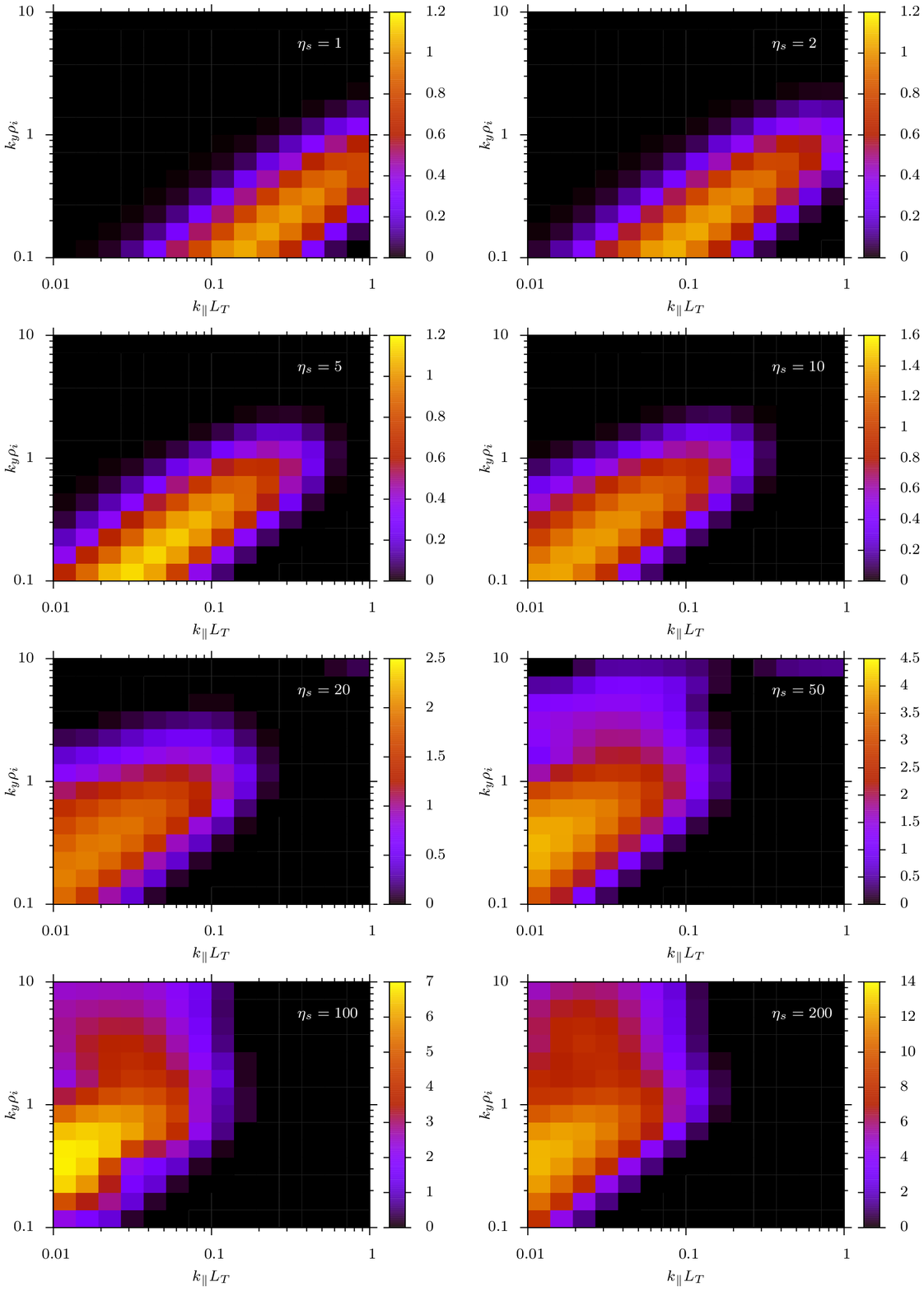,height=8in}}
\caption{\label{fig:N2D} The amplification exponent $N$ vs.\ $\kpar\LT$ 
and $k_y\rhoi$, obtained numerically using {\tt AstroGK} for the same  
parameters as the curves in the left panel of \figref{fig:N}.} 
\end{figure}

\subsection{Case of finite shear}
\label{sec:finiteS}

In the intermediate regimes between large and small $\chi$ 
(i.e., weak and strong shear), the solutions of \eqref{eq:itgt}
transit from the weak-shear form described in \secref{sec:weakS} 
to the pure-PVG case treated in \secref{sec:longt}. 
\Figref{fig:gamma_itg} shows the time-dependent growth rates 
and frequencies for several values of $\chi$. As $\chi$ decreases 
(i.e., $S$ increases), the peak of the growth rate moves 
further into the past and the growth rate asymptotes 
to the pure-PVG case (\figref{fig:gamma}). 
It is not hard to convince oneself analytically that this is indeed 
what ought to happen. Since intuitively it is rather obvious, 
further asymptotic considerations on this subject are exiled to \apref{app:longt}. 
The long-time decay asymptotic is also derived there (it is exactly 
analogous to that found in \secref{sec:decay}).

Similarly to our previous calculations, the amplification exponent is 
\beq
N(\zS) 
= t_0(\zS)\sqrt{\zS^2+1}\int_0^1 d\xi\,\Im\,\tz(\xi,\chi(\zS)),
\label{eq:N_itg}
\eeq
where $\tz$ is the solution of \eqref{eq:itgt}. 
The wavenumber dependence enters via the $\zS$ dependence 
of $t_0$ and of $\chi$ (see \eqref{eq:chi}).
The numerically computed amplification exponent as a function 
of $\zS$ and of $\eS$ is plotted in the left panel of \figref{fig:N} 
for $\eta_i=5$ and $q/\ee=10$ 
(these are representative of the values encountered 
in the more realistic numerical studies of tokamak 
transport \cite{Barnes11,Highcock10,Highcock11}). 
The pure-PVG case treated in \secref{sec:longt} remains 
a good approximation up to values of $\eS$ of order $10$. 
After that, there is a transition towards the weak-shear limit 
(\secref{sec:weakS}), accompanied by the 
loss of wavenumber dependence as $\eS$ 
is increased to values of order $100$. 
The constant of proportionality between $k_y$ and $\kpar$ 
for the maximally amplified modes (i.e., $\zS$ at which $N$ is 
maximised) does not appear to depend on $\eS$, although 
at large $\eS$, the maximum is increasingly weak.  
The maximal amplification exponent is plotted in the 
right panel of \figref{fig:N}. It is perhaps worth pointing 
out the qualitative similarity between this plot and figure~1 of 
Ref.~\cite{Highcock10}, obtained from gyrokinetic simulations 
in full tokamak geometry.  

Finally, the amplification exponent as a function of $k_y$ and $\kpar$, 
obtained in direct (linear) numerical simulations, is shown in \figref{fig:N2D}
(the parameters are the same as in the ``theoretical'' \figref{fig:N}, left panel).  
The transition from the PVG curve \eqref{eq:Nmax} to the flat wavenumber 
dependence \eqref{eq:N_weakS} is manifest, as are the limits of 
applicability of our approximations in the wavenumber space. Note 
the different normalisation of the parallel wavenumber here ($\kpar\LT$, characteristic 
of ITG) compared to the left panel of \figref{fig:kspace} ($\kpar\vthi/S$, characteristic 
of PVG). Hence the drift towards higher $\kpar\LT$ as $\eS=\vthi/S\LT$ decreases 
towards the PVG-dominated regime, where the parallel scale of maximally amplified 
modes is set by the shear rather than the temperature gradient.



\section{Qualitative summary of the linear results}
\label{sec:qualit} 

In a gyrokinetic plasma with radial gradients of temperature and parallel velocity, 
both gradients are sources of free energy and so will drive the growth of fluctuations 
(ITG and PVG instabilities). The typical growth rate is of order $\gamma\sim \vth/\LT$ for ITG  
and $\gamma\sim qS/\ee$ for PVG (see \eqref{eq:gamma_gen_pvg}), 
or the mean square of the two if they are comparable (see \eqref{eq:omega_gen}). 
Because the mean plasma velocity is toroidal, it always has both a parallel and 
a perpendicular component (the latter a factor of $q/\ee$ smaller than the former). 
The shear in the perpendicular ($\vE\times\vB$) velocity is stabilising and causes 
all modes to decay eventually, so the fluctuation growth is transient --- it is always transient 
in the limit, considered here, of zero magnetic shear and it is transient for large enough 
velocity shear $S$ when the magnetic shear is finite \cite{Newton10,Barnes11}. 
If the linear physics provides sufficiently vigorous and lasting amplification of finite initial 
perturbations, it is intuitively clear that 
the system must be able to sustain nonlinearly a saturated (subcritical) turbulent state
(see \secref{sec:turb}). Therefore, the interesting question is how much transient amplification 
should be expected to occur and on what time scale. 

In the preceding sections, we have addressed this question mathematically, with the 
results summarised by \figsref{fig:kspace}, \ref{fig:N} and \ref{fig:N2D} 
(see also \eqref{eq:Nmax}, \eqref{eq:t0_max} and \eqref{eq:t0_weakS_dim}). 
Very roughly, these results can be explained as follows. The effect of the perpendicular shear 
is to produce a secular increase with time of the radial wavenumber, 
$k_x(t)\sim Sk_y t$. When this becomes large enough, the instability is killed by 
Landau damping (see discussion at the end of \apref{app:pvg}). 
If we estimate that this happens after $t_0\sim S^{-1}$ (i.e., 
for $k_x(t_0)\rhoi\sim1$, assuming $k_y\rhoi\sim1$), 
we may conclude that initial perturbations will be amplified by a factor 
of $e^N$, where the amplification exponent is 
\beq
N\sim\gamma t_0 \sim \frac{\vthi}{\LT S}~~{\rm for~ITG\quad and}\quad  
N\sim \frac{q}{\ee}~~{\rm for~PVG}.
\eeq
Thus, the shear quenches the ITG amplification 
--- but $N$ cannot fall below the shear-independent level associated with the PVG (\figref{fig:N}, right panel). 
This is indeed the case (see \eqref{eq:Nmax} and \eqref{eq:t0_weakS_dim}), 
although, strictly speaking, one has to take into account the dependence of the quenching effect 
on the perpendicular and parallel wavenumbers --- long-wavelength modes grow more slowly, 
but for a longer time; in the case of PVG, there is also a preferred relationship 
$k_y\rhoi\sim (\ee/q)^{1/3}\kpar\vthi/S$ for the most strongly amplified modes 
(see \secref{sec:longt} and \figref{fig:N2D}). While these wavenumber dependences 
are likely to be important in the analysis of the resulting turbulent state and the associated 
transport, they effectively cancel out in the expression for the amplification exponent
(because $\gamma\propto k_y\rhoi$, $t_0\propto 1/k_y\rhoi$ at long wavelengths) 
and the results of the qualitative argument that we have given hold true.  

It is instructive to compare these results with the conclusions of a long-wavelength 
fluid ITG-PVG theory presented in \cite{Newton10} (for the case of finite magnetic shear). 
In that regime, perpendicular shear, by effectively increasing $k_x(t)$, also caused eventual damping of 
the fluctuations, but this time via collisional viscosity. Therefore, to estimate the 
transient growth time $t_0$, one must set $\gamma \sim \nuii k_x^2(t_0)\rhoi^2 
\sim \nuii S^2k_y^2\rhoi^2t_0^2$, where $\nuii$ is the ion collision rate. 
Then, ignoring wavenumber dependences again,
$t_0\propto\gamma^{1/2}S^{-1}$, so the amplification exponent is 
\beq
N\sim\gamma t_0\propto\frac{\gamma^{3/2}}{S}
\propto\frac{1}{S}~~{\rm for~ITG\quad and}\quad  
N \propto \sqrt{S}~~{\rm for~PVG}.
\eeq
Thus, the $\vE\times\vB$ velocity shear again quenches the ITG instability, 
but once $S$ is large enough for the PVG drive to take over, the amplification exponent
actually grows as $\sqrt{S}$, the result obtained rigorously by \cite{Newton10} --- in contrast 
with the shear-independent $N\sim q/\ee$ that we have found in the kinetic regime. 
The practical conclusion from this is that it should be easier 
to obtain states of reduced transport \cite{Barnes11,Highcock10,Highcock11,Parra11} 
in weakly collisional, kinetic plasmas. 

\section{Subcritical PVG turbulence and transport} 
\label{sec:turb}

While a detailed nonlinear theory is not the primary objective of this paper, 
we would like to give a preliminary, very crude and very heuristic discussion, 
inspired by the linear results presented above and by previous work on turbulence and 
transport scalings \cite{Barnes11itg}. 

\subsection{Turbulence threshold}
\label{sec:marginal}

Consider a situation when initial perturbations can grow transiently. 
The reason they decay eventually is that their radial wavenumber $k_x(t)$ gets 
swept by the $\vE\times\vB$ shear from the unstable to damped region
(see discussion at the end of \apref{app:pvg} and at the end of \apref{app:itg}). 
If nonlinear interactions can scatter the energy from these modes 
back into the unstable region before they decay to small amplitudes 
beyond the reach of nonlinearity, then they can be transiently amplified once 
again and so on. Thus a nonlinear saturated state can be sustained
--- the subcritical PVG turbulence. This argument is entirely analogous 
to the standard paradigm for subcritical turbulence in hydrodynamic 
shear flows \cite{Trefethen93}. 

The typical time scale for the nonlinear 
interactions is the nonlinear decorrelation time $\sim 1/\kperp\du$, where 
$\du \sim \kperp(c\ephi/B)$ is the fluctuating $\vE\times\vB$ velocity. 
Therefore, in order to sustain turbulence the transient growth should 
last at least as long as one nonlinear decorrelation time: 
\beq
\label{eq:t0_crit}
t_0 \gtrsim \frac{1}{\kperp\du},
\eeq
where $t_0$ is the amplification time. In the saturated state, if it is sustained, 
the rate of amplification should be comparable to nonlinear decorrelation rate: 
\beq
\geff \sim \frac{N}{t_0} \sim \kperp\du,
\label{eq:geff}
\eeq
where $N$ is the amplification exponent. 
Combined with \eqref{eq:t0_crit}, this immediately implies, unsurprisingly, 
that the criterion for onset of turbulence is
\beq
N\gtrsim 1.
\eeq  
For subcritical turbulence, this criterion replaces the marginal stability condition $\gamma=0$, 
usually employed for cases with well-defined eigenmodes. 
It is the ``significant amplification threshold'' introduced in \secref{sec:threshold}. 
As we saw there, for PVG turbulence in a slab, it is equivalent to $q/\ee\gtrsim7$. 
One might expect that much more sophisticated criteria could be derived 
by refining our arguments and testing these refinements against dedicated 
numerical parameter scans. The key conclusion is that $q/\ee$ is now the critical 
parameter to be tuned and so magnetic configurations in which it is 
smaller may hold the promise of reduced or even completely 
suppressed ion turbulent transport (e.g., spherical tokamaks \cite{Roach09}). 

\subsection{Transport scalings}
\label{sec:scalings}

The standard mixing-length heuristics (often, somewhat misleadingly, referred 
to as ``quasilinear'' theory) are based on the argument that if fluctuations are 
driven by a linear instability with a characteristic growth rate $\gamma$, then 
they will saturate at amplitudes and scales such that the nonlinear decorrelation 
rate is comparable to this growth rate, viz., $\kperp \du\sim \gamma$. 
Then the turbulent heat diffusivity is $\chiT\sim\du/\kperp\sim\gamma/\kperp^2$ 
and so the ion heat flux is $Q_i\sim \chiT n_iT_i/\LT$. 

For subcritical fluctuations, there is no definite $\gamma$, but it is intuitive 
to argue, as we did in \secref{sec:turb}, that it should be replaced by $\geff\sim N/t_0$. 
For the PVG-driven fluctuations, we showed 
in \secref{sec:Nmax} that, at maximal amplification, 
$N\sim q/\ee$ and $t_0\sim (q/\ee)/\kpar\vthi$, so \eqref{eq:geff} gives
\beq
\geff\sim\kpar\vthi \sim \kperp\du.
\eeq 
This simple estimate is actually consistent with a very general 
idea that in systems with parallel propagation (or particle streaming) and perpendicular 
nonlinearity, turbulence tunes itself into a {\em critically balanced} state, viz., the time scales 
for these two effects are always comparable \cite{Goldreich95,Cho04,Schekochihin08,Tome,Nazarenko11,Barnes11itg}. 
We may now use the relationship \eqref{eq:Nmax} between $\kpar$ and $k_y$ for the maximally 
amplified modes to estimate $\kperp\rhoi\sim (\ee/q)^{1/3}\kpar\vthi/S$ and conclude, therefore, 
that 
\beq
\chiT\sim \frac{\geff}{\kperp^2} \sim 
\rhoi^2\,\frac{S^2}{\kpar\vthi}\lt(\frac{q}{\ee}\rt)^{2/3}
\sim \frac{\rhoi^2 R}{\vthi}\frac{S^2 q^{5/3}}{\ee^{2/3}}. 
\label{eq:chiT}
\eeq 
In the last expression, we have made another important assumption: 
since it is the longest-wavelength fluctuations that dominate 
transport, we should use the lowest parallel wavenumber possible 
in a tokamak: $\kpar\sim1/qR$, where $R$ is the major radius. 
This prescription was proposed in \cite{Barnes11itg} for 
ITG turbulence. In a sense, it is even more natural here than it was 
there because, in the theory developed in the preceding sections, the wavelengths 
of maximally amplified modes are not limited from above by any microscale 
physics --- if we took the slab model literally, the limit would be the 
periodicity length of the box; in a tokamak, the connection length is a 
natural choice. Finally, under this scheme, the ion heat flux scales 
as\footnote{There is a number of reasons to take these specific 
scaling predictions with a grain of salt. 
Besides making the assumptions stated above, we have ignored many effects 
that may be important and may change our estimates of the relevant scales, 
times and amplitudes: the role of zonal flows in regulating and/or sustaining 
the turbulence, the possibility that 
the effective radial and poloidal wavenumbers, $k_x$ and $k_y$, are not 
the same in a system with imposed velocity shear, the role of magnetic shear 
if it is present, various geometry (curvature) effects etc. 
It is also not necessary, although intuitive and possibly 
supported by numerical evidence \cite{Highcock11}, that the relationship 
\eqref{eq:Nmax} between $\kpar$ and $k_y$ obtained by maximising the linear 
amplification should persist in the nonlinear regime, especially if the system 
is far above the significant amplification threshold (\secref{sec:marginal}).  
There are several possible alternative theories that can be constructed 
in a similar vein to that presented above, but the current state of numerical 
and experimental evidence does not yet allow us to differentiate between them 
in a falsifiable fashion. Future investigations will focus on this task. Preliminary 
numerical studies suggest that far from the significant amplification 
threshold (\secref{sec:marginal}), the ion heat flux does not in fact scale as strongly 
with $q$ as suggested by \eqref{eq:Qi} (E~G~Highcock 2011, unpublished). }  
\beq
\frac{Q_i}{n_i T_i\vthi} \sim \frac{\chiT}{\vthi\LT} \sim 
\lt(\frac{S}{\Omega_i}\rt)^2 \frac{q^{5/3}}{\ee^{2/3}}\frac{R}{\LT}. 
\label{eq:Qi}
\eeq  

One fairly obvious feature of the scalings \eqref{eq:chiT} and \eqref{eq:Qi} 
(independent of most of the specific 
assumptions that we made in deriving them) is that the heat diffusivity is independent 
of the temperature gradient and so heat transport is not very ``stiff'' --- the relevant 
comparison is with the scaling for the ITG regime, $Q_i\propto q (R/\LT)^3$ \cite{Barnes11itg}. 
A softening of transport in the presence of velocity shear has indeed been reported 
both in experimental \cite{Mantica09,Mantica11} and 
in numerical \cite{Barnes11,Highcock10,Highcock11} studies. Physically, it is 
not surprising: as the driver of the turbulence in this regime is the PVG, not the ITG, 
steeper temperature gradients do not produce stronger turbulence and so the 
positive feedback loop between $R/\LT$ and the heat diffusivity is broken.

\ack 
We gratefully acknowledge many inspiring discussions with M~Barnes, J~Connor,  
N~Loureiro, F~Parra, C~Roach and especially W~Dorland.
Some of these interactions were made possible by the Leverhulme Trust International 
Network for Magnetised Plasma Turbulence.
Some of the work reported herein was done at the Isaac Newton Institute, Cambridge, during the 
programme ``Gyrokinetics in Laboratory and Astrophysical Plasmas'' (2010). 
Numerical simulations were carried out at HPC-FF (J\"ulich) and HECTOR (Edinburgh). 
AAS was supported in part by the STFC Grant ST/F002505/2. 
EGH was supported by an EPSRC CASE studentship in partnership with the 
Euratom/CCFE Association.
The views and opinions expressed herein are unlikely to reflect 
those of the European Commissioners.  
EGH was also supported in part by the Thematic Programme ``Gyrokinetics for ITER''
at the Wolfgang Pauli Institute, Vienna (2011). 

\appendix

\section{Low-Mach-number local gyrokinetics in a rotating 
axisymmetric plasma}
\label{app:gk}

Here we describe the version of the gyrokinetic system of equations appropriate 
for a rotating axisymmetric plasma, which is the starting point for our calculation. 
For a detailed derivation, we refer 
the reader to \cite{Abel11} (earlier treatments are \cite{Artun94,Sugama97,Peeters09}). 

We consider the axisymmetric rotating equilibrium and work in the 
subsidiary low-Mach-number limit, as described at the beginning 
of \secref{sec:gk}. 
The distribution function of particles of species $s$ 
is written in the following form
\beq
\fl
f_s(\vr,\vv) = \lt[1-\frac{Z_s e \dphi(\vr)}{T_s}\rt]F_{0s}(\psi(\vR_s),\eps_s) 
+ F_{1s}(\vR_s,\eps_s,\mu_s,\sigmapar) + h_s(\vR_s,\eps_s,\mu_s,\sigmapar), 
\label{eq:fs}
\eeq 
Let us explain the numerous notation that appears here. 
The standard 6D kinetic position-and-velocity phase space $(\vr,\vv)$ 
is transformed to the 5D gyrokinetic phase space, where the 
dynamics are averaged over the Larmor orbits and so do not depend 
on the gyroangle.
If $\vw = \vv-\vu$ is peculiar velocity with respect to the mean flow,
$B$ is the magnitude of the mean magnetic field, $\vb=\vB/B$ its direction, 
$\Omega_s=Z_seB/m_s c$ the cyclotron frequency, $Z_s e$ is particle charge
($Z_e=-1$), $m_s$ particle mass, $c$ the speed of light, 
then the gyrokinetic variables $(\vR_s,\eps_s,\mu_s,\sigmapar)$ 
are defined as follows: 
the guiding centre position $\vR_s = \vr - \vb\times\vw/\Omega_s$, 
the energy variable $\eps_s = m_sw^2/2$ (this is only correct in the low-$M$ limit), 
the magnetic moment $\mu_s=m_s\wperp^2/2B$, 
and the sign of the parallel velocity $\sigmapar=\wpar/|\wpar|$
(the subscripts $\perp$ and $\parallel$ refer to the mean-field direction $\vb$). 
In \eqref{eq:fs}, the particle distribution function is split into the 
mean Maxwellian, which can be shown to depend only on the 
flux label $\psi(\vR_s)$ via the mean density $n_s$ and 
mean temperature~$T_s$, namely,\footnote{Establishing the Maxwellian equilibrium depends 
on the plasma being sufficiently 
collisional, namely, that the collision frequency is not smaller  
than the fluctuation frequency by more than one order in 
the gyrokinetic expansion parameter. 
The density being a function solely of $\psi$ only holds in the low-$M$ limit.} 
\beq
F_{0s}(\psi(\vR_s),\eps_s) = n_s(\psi(\vR_s))
\lt[\frac{m_s}{2\pi T_s(\psi(\vR_s))}\rt]^{3/2}
\exp\lt[-\frac{\eps_s}{T_s(\psi(\vR_s))}\rt],
\eeq 
the mean perturbed distribution function $F_{1s}$, which contains 
collisional (classical and neoclassical) effects and will not concern us 
here, the Boltzmann response containing the perturbed scalar potential 
$\dphi$, and the guiding-centre distribution function $h_s$. 

If we take the mean fields $\vB$ (i.e., $\psi$ and $I(\psi)$), 
$n_s(\psi)$, $T_s(\psi)$ and $\omega(\psi)$ 
to be known and the fluctuations about them to be purely electrostatic, 
then the latter are fully described by a closed system containing the 
gyrokinetic evolution equation for $h_s$ and the quasineutrality condition 
determining $\dphi$: 
\bea
\fl
\nonumber
\frac{\dd h_s}{\dd t} 
+ [\vu(\vR_s) + \wpar\vb + \vV_d + \avgs{\vV_E}]\cdot\frac{\dd h_s}{\dd\vR_s}
= \frac{Z_se F_{0s}}{T_s}\lt[\frac{\dd}{\dd t} 
+ \vu(\vR_s)\cdot\frac{\dd}{\dd\vR_s}\rt]\avgs{\dphi}\\
\label{eq:gk_gen}
- (\avgs{\vV_E}\cdot\vdel\psi)\lt[\frac{d\ln n_s}{d\psi} + 
\lt(\frac{\eps_s}{T_s} - \frac{3}{2}\rt)\frac{d\ln T_s}{d\psi} + 
\frac{B_\phi}{B}\frac{m_s\wpar R}{T_s}\frac{d\omega}{d\psi}\rt]F_{0s},\\
\lt(\sum_s\frac{Z_s^2e^2 n_s}{T_s}\rt)\dphi = \sum_s Z_s e\int d^3\vw\avgr{h_s},
\label{eq:quasineut_gen}
\eea
where $\wpar=\sigmapar\sqrt{2(\eps_s-\mu_sB)}$, 
$B_\phi=I(\psi)/R$ is the azimuthal magnetic field, 
$\vV_d = (c/Z_s eB)\vb\times(m_s\wpar^2\vb\cdot\vdel\vb + \mu_s\vdel B)$
is the magnetic drift velocity, 
$\vV_E = (c/B)\vb\times\vdel\dphi$ is the $\vE\times\vB$ velocity,
and the gyroaverages are defined 
\bea
\fl
\avgs{\dphi} = \frac{1}{2\pi}\int_0^{2\pi} d\vartheta\, 
\dphi(\vR_s+{\vb\times\vw}/{\Omega_s})\quad \rm{(function~of}~\vR_s\rm{)},\\
\fl
\avgr{h_s} = \frac{1}{2\pi}\int_0^{2\pi} d\vartheta\, 
h_s(\vr-{\vb\times\vw}/{\Omega_s},\eps_s,\mu_s,\sigmapar)\quad 
\rm{(function~of}~\vr~\rm{and}~\vw\rm{)}.
\eea
Note that we suppressed the collision term in \eqref{eq:gk_gen}
(formally, we are ordering the collision frequency small via a subsidiary expansion). 

The approximation of Boltzmann electrons amounts to setting $h_e=0$ 
in \eqref{eq:quasineut_gen}. 

\subsection{Local Cartesian frame in the case of zero magnetic shear}
\label{app:shbox}

Let us take the mean magnetic field to be locally straight and uniform, 
so it has constant magnitude, no curvature and no shear. 
This means that the magnetic drifts vanish ($\vV_d=0$) 
and we can introduce a local orthogonal Cartesian coordinate system 
(see \figref{fig:xyz}):
\beq
\vx=\frac{\vdel\psi}{B_p R},\quad
\vy=\frac{\vb\times\vdel\psi}{B_p R},\quad
\vz=\vb,
\eeq
where $B_p=|\vdel\psi|/R$ is the poloidal component of $\vB$. 
We will view this coordinate system as having its origin ($x=0$) at 
some reference flux surface $\psi_0$, so in the vicinity 
of this flux surface, $\psi \approx \psi_0 + xB_p R$. 
Then 
\beq
\fl
\vu = \omega(\psi)R^2\vdel\psi = 
\omega(\psi)R\lt(\frac{B_\phi}{B}\,\vz + \frac{B_p}{B}\,\vy\rt)
\approx \lt[\omega R + x B_p R^2\frac{d\omega}{d\psi}\rt]
\lt(\frac{B_\phi}{B}\,\vz + \frac{B_p}{B}\,\vy\rt),
\eeq
where all quantities in the last expression are taken at $\psi=\psi_0$
and $B_\phi=I(\psi)/R$ is the azimuthal component of $\vB$. 
We have only retained the spatial dependence  
of $\omega$ (in the form of a Taylor expansion) because we are formally 
ordering velocity gradients as $O(1/M)$, while all other mean fields 
are assumed to have $O(1)$ variation on the system scale. 
The constant part of the velocity can be removed by going to a frame 
moving (azimuthally) at this velocity. The $z$ component of the velocity 
shear can be neglected because it multiplies $\dd/\dd z$ in \eqref{eq:gk_gen}
and we order $x$ as small (in the gyrokinetic expansion parameter) compared 
to the parallel scale of the fluctuations, whereas the $y$ component of the 
shear remains important because it multiplies $\dd/\dd y\gg\dd/\dd z$. 
It follows that the velocity field 
can be replaced by the linear shear flow as stated in \eqref{eq:Sdef}. 

Given the shearing box approximation described above and the assumption 
of Boltzmann electrons ($h_e=0$), 
the conversion of \eqsand{eq:gk_gen}{eq:quasineut_gen} 
into \eqsand{eq:gk_sh}{eq:quasineut_sh} is straightforward. 
Note that $(c/B)(T_i/Ze)=\vthi^2/2\Omega_i=\rhoi\vthi/2$,
where $\rhoi=\vthi/\Omega_i$ is the ion Larmor radius.

\section{PVG and ITG-PVG dispersion relations}

Here we provide some elementary analytical considerations of the linear 
dispersion relations that govern the early evolution of the fluctuations. 

\subsection{PVG dispersion relation}
\label{app:pvg}

In the pure PVG case, the dispersion relation is \eqref{eq:pvg}, containing 
the well-known PVG instability \cite{Catto73} and the sound wave. 
The following considerations provide some analytical support for \figref{fig:pvg}. 

\paragraph{Marginal stability threshold.}

This is easy to obtain analytically without 
any approximation: at the threshold, $\Im\,\zz=0$ and then, from \eqref{eq:pvg}, 
the real frequency is readily shown to vanish as well, so we get 
$(q/\ee)\zS = (1+\tau/Z)/\Gamma_0(\lambda)$. 
Thus, we have an instability if\,\footnote{If $qS/\ee>0$, the unstable 
mode has either $k_y>0$, $\kpar>0$ or $k_y<0$, $\kpar<0$ (see \figref{fig:itg}
and discussion in \apref{app:itg}); 
if $qS/\ee<0$, then $k_y$ and $\kpar$ have to have opposite signs. We can assume 
without loss of generality that all these quantities are positive.} 
\beq
\frac{\kpar\vth}{qS/\ee} < \frac{k_y\rhoi\Gamma_0(\lambda)}{1+\tau/Z}
\lesssim \frac{0.66}{1+\tau/Z},
\label{eq:marg}
\eeq 
where $\lambda=k_y^2\rhoi^2/2$ (see \figref{fig:pvg}, right panel).
The second inequality in \eqref{eq:marg} implies that there is an absolute
finite limit on the parallel wavenumbers at which the PVG instability can 
survive (reached for $k_y\rhoi\approx1.26$).

\paragraph{PVG modes and sound waves.}

The growing modes have no real frequency. In fact, there is no real frequency
as long as $(q/\ee)\zS>1$, which means that the mode is purely decaying
between $(q/\ee)\zS=1$ and the marginal stability boundary 
$(q/\ee)\zS=(1+\tau/Z)/\Gamma_0(\lambda)>1$ (see \eqref{eq:marg}). 
For $(q/\ee)\zS<1$, the mode turns into a damped sound wave (this is shown in 
the left panel of \figref{fig:pvg}). 
Thus, the PVG drive has pushed the sound waves into a wedge 
in wavenumber space, 
\beq
k_y\rhoi<\frac{\kpar\vth}{qS/\ee} 
\label{eq:soundwedge}
\eeq
(see \figref{fig:pvg}, right panel) 
and populated the rest with nonpropagating growing or decaying modes.

\paragraph{Growth at long parallel wavelengths.}

Looking for unstable solutions in the long-wavelength limit 
($\kpar\vth/S\ll1$), we consider 
the asymptotic form of \eqref{eq:pvg} with $\zz=i\gbar$ (pure growth) 
and $\gbar\gg1$, when $1+i\gbar \Zfn(i\gbar)\approx 1/2\gbar^2$.
Then the growing solution~is
\beq
\fl
\gbar \approx
\sqrt{\frac{(q/\ee)\zS-1}{2[(1+\tau/Z)/\Gamma_0(\lambda) - 1]}}
\quad\Rightarrow\quad
\gamma \approx \sqrt{\frac{(qS/\ee)k_y\rhoi\kpar\vth}{2[(1+\tau/Z)/\Gamma_0(\lambda) - 1]}}.
\label{eq:gamma_long}
\eeq
In the second expression, we have restored dimensions and assumed 
$(q/\ee)\zS\gg1$ (which ensures $\gbar\gg1$ and is consistent with 
the short-time ordering adopted at the beginning 
of \secref{sec:pvg}). 
Note that, while this asymptotic has a peak at $k_y\rhoi\sim1$,  
it does not capture the maximum growth rate because of its monotonic 
increase with $\kpar$. The maximum growth rate is, in fact, 
reached for $\gbar\sim1$, where the plasma dispersion function does 
not yield itself to a simple asymptotic expansion. The numerical 
solution is shown in \figref{fig:pvg}.

\paragraph{Growth at short perpendicular wavelengths.}

Finally, we note that one can obtain a good approximate preview of the 
results of \secref{sec:longt} if one restores the dependence on $k_x$ 
by setting $\lambda = (k_x^2+k_y^2)/2$ in \eqref{eq:pvg}. The effect 
of the perpendicular shear is to increase the instantaneous $k_x$ of the mode, 
so the limit of short perpendicular wavelengths (or, more 
precisely, large $k_x$ but finite $k_y$) is similar to the limit of long times. 
Since $\Gamma_0(\lambda)\approx1/\sqrt{2\pi\lambda}$ for $\lambda\gg1$, 
\eqref{eq:pvg} becomes (assuming also $k_y\rhoi\gg\kpar\vth/(qS/\ee)$, 
i.e., we are far outside the domain of the damped sound waves discussed above): 
\beq 
\lt(1+\frac{\tau}{Z}\rt)\sqrt{\pi}\,\frac{\kpar\vth}{qS/\ee} \sqrt{1+\frac{k_x^2}{k_y^2}} 
\approx \frac{1}{St_0} \sqrt{1+\frac{k_x^2}{k_y^2}} 
\approx 1+i\gbar \Zfn(i\gbar),
\label{eq:large_kperp}
\eeq
where $t_0$ is given by \eqref{eq:t0}, but with dimensions restored 
(i.e., not normalised by $\kpar\vthi$) and $\ee/q\ll\zS\ll1$ (as will 
be the case for the most strongly amplified modes in \secref{sec:Nmax}). 
This formula has two interesting consequences. 

Firstly, setting $k_x=Sk_yt$ and $St\ll1$, we recover the time-dependent 
dispersion relation \eqref{eq:pvgt} and so the transient growth and 
eventual decay of the fluctuations derived more formally in \secref{sec:longt}
(without the assumption $\ee/q\ll\zS\ll1$). 
Thus, there is a smooth connection between the short-time and long-time behaviour
of the PVG-driven fluctuations. Note that applying the same asymptotics to 
\eqref{eq:gamma_long}, we recover~\eqref{eq:growth}.  
That the PVG growth rate must be extinguished at large enough $k_x$ is 
also obvious from the instability criterion \eqref{eq:marg}, where 
increasing $\lambda$ eventually breaks the first inequality. 

Secondly, in the short-time but also short-perpendicular-wavelength 
limit ($k_x\ll k_y$ but $k_y\rhoi\gg1$), 
the growth rate of the PVG instability is independent of $k_y$, 
as is indeed manifest in \figref{fig:pvg}. 

\subsection{ITG-PVG dispersion relation}
\label{app:itg}

In the short-time limit, 
we can, analogously to the derivation in \secref{sec:pvg}, 
reduce the integral equation \eqref{eq:inteq} to a dispersion 
relation for the normalised complex frequency $\zz=\omega/|\kpar|\vth$: 
\beq
\fl
\frac{1+\tau/Z}{\Gamma_0(\lambda)} - 1 = 
\lt(\frac{q}{\ee}\,\zS-1-\eta_i\zn\zz\rt)[1+\zz\Zfn(\zz)]
+ \eta_i\zn\lt[\frac{3}{2}-\Lambda(\lambda)-\frac{1}{\eta_i}\rt]\Zfn(\zz),
\label{eq:disp_rln_itg}
\eeq
where $\Gamma_0(\lambda) = e^{-\lambda} I_0(\lambda)$ and 
$\Lambda(\lambda)=1-\lambda + \lambda I_1(\lambda)/I_0(\lambda)$; 
we recapitulate the definitions of 
$\lambda = k_y^2\rhoi^2/2$, 
$\zS = S k_y\rhoi/\kpar\vth$, 
$\zn = k_y\rhoi/2|\kpar|\Ln$ 
and $\eta_i=\Ln/\LT$.  
This is the standard slab ITG-PVG dispersion relation for an 
unsheared gyrokinetic plasma. 

\paragraph{Marginal stability thresholds.} The way in which ITG and PVG coexist is 
easiest to understand by examining the marginal stability thresholds. 
Setting $\Im\,\zz = 0$, $\zz = \Re\,\zz = \zz_0$ and demanding 
that the imaginary part of \eqref{eq:disp_rln_itg} vanish, 
we get the equation for the real frequency of the mode at marginal stability: 
\beq
\eta_i\zn\zz_0^2 - \lt(\frac{q}{\ee}\,\zS-1\rt)\zz_0 - 
\eta_i\zn\lt[\frac{3}{2} - \Lambda(\lambda) - \frac{1}{\eta_i}\rt] = 0.
\eeq
Substituting the solution of this equation back into \eqref{eq:disp_rln_itg}, 
we arrive at the marginal stability condition: 
\beq
\eta_i^2\zn^2 \lt[\frac{3}{2} - \Lambda(\lambda)-\frac{1}{\eta_i}\rt]
+ \lt[\frac{1+\tau/Z}{\Gamma_0(\lambda)}-1\rt]\lt[\frac{q}{\ee}\,\zS
- \frac{1+\tau/Z}{\Gamma_0(\lambda)}\rt] = 0.
\label{eq:marg_eqn}
\eeq
Since $\eta_i\zn=k_y\rhoi/2|\kpar|\LT$ and $\zS=Sk_y\rhoi/\kpar\vthi$, the above 
equation can be solved for the two curves in the $(\kpar,k_y)$ plane 
that enclose the unstable region: 
\beq
\fl
\kpar\LuT = \frac{k_y\rhoi\,\eta_u^2}{2\sqrt{1+\eta_u^2}}
\frac{\lt[3/2 - \Lambda(\lambda) - 1/\eta_i\rt]/\lt[(1+\tau/Z)/\Gamma_0(\lambda)-1\rt]}{\lt\{-1\pm
\sqrt{1+\eta_u^2\lt[3/2 - \Lambda(\lambda) - 1/\eta_i\rt]/\lt[1-\Gamma_0(\lambda)/(1+\tau/Z)\rt]}\rt\}},
\label{eq:marg_gen}
\eeq
where $\LuT=\vthi/\sqrt{(qS/\ee)^2+(\vth/\LT)^2} = (qS/\ee)\sqrt{1+\eta_u^2}$ is 
a convenient normalisation of the parallel wavenumber for the mixed ITG-PVG regime
and $\eta_u=(\vthi/\LT)/(qS/\ee)=\eS/(q/\ee)$ is a parameter that measures the relative 
strength of the ITG and PVG drives. 

The pure PVG regime is $\eta_u\ll1$, in which case the ``$+$'' curve above 
turns into \eqref{eq:marg} (which is perhaps easier to infer directly from 
\eqref{eq:marg_eqn}), while the ``$-$'' curve is simply $\kpar=0$ --- these 
demarcate two symmetric PVG-unstable regions 
at $\kpar>0$, $k_y>0$ and $\kpar<0$, $k_y<0$ (assuming $S>0$). 
The pure ITG regime is $\eta_u\gg1$, in which case $\LuT=\LT$ and  
\beq
\kpar\LT = \pm k_y\rhoi\sqrt{
\frac{3/2 - \Lambda(\lambda) - 1/\eta_i}{4\lt[(1+\tau/Z)/\Gamma_0(\lambda)-1/2\rt]^2-1}}.
\label{eq:marg_itg}
\eeq
These curves enclose four symmetric ITG-unstable regions in the four 
quadrants of the $(\kpar,k_y)$ plane. In the general case of finite $\eta_u$, when 
both ITG and PVG drives play a role, these drives combine constructively in the 
PVG-unstable quadrants $\kpar>0$, $k_y>0$ and $\kpar<0$, $k_y<0$, whereas in the 
remaining two quadrants, the stable PVG mode, which is just a PVG-modified sound 
wave (cf.~\eqref{eq:sound_wave}), has a stabilising influence on the ITG drive.   

\begin{figure}[t]
\centerline{\epsfig{file=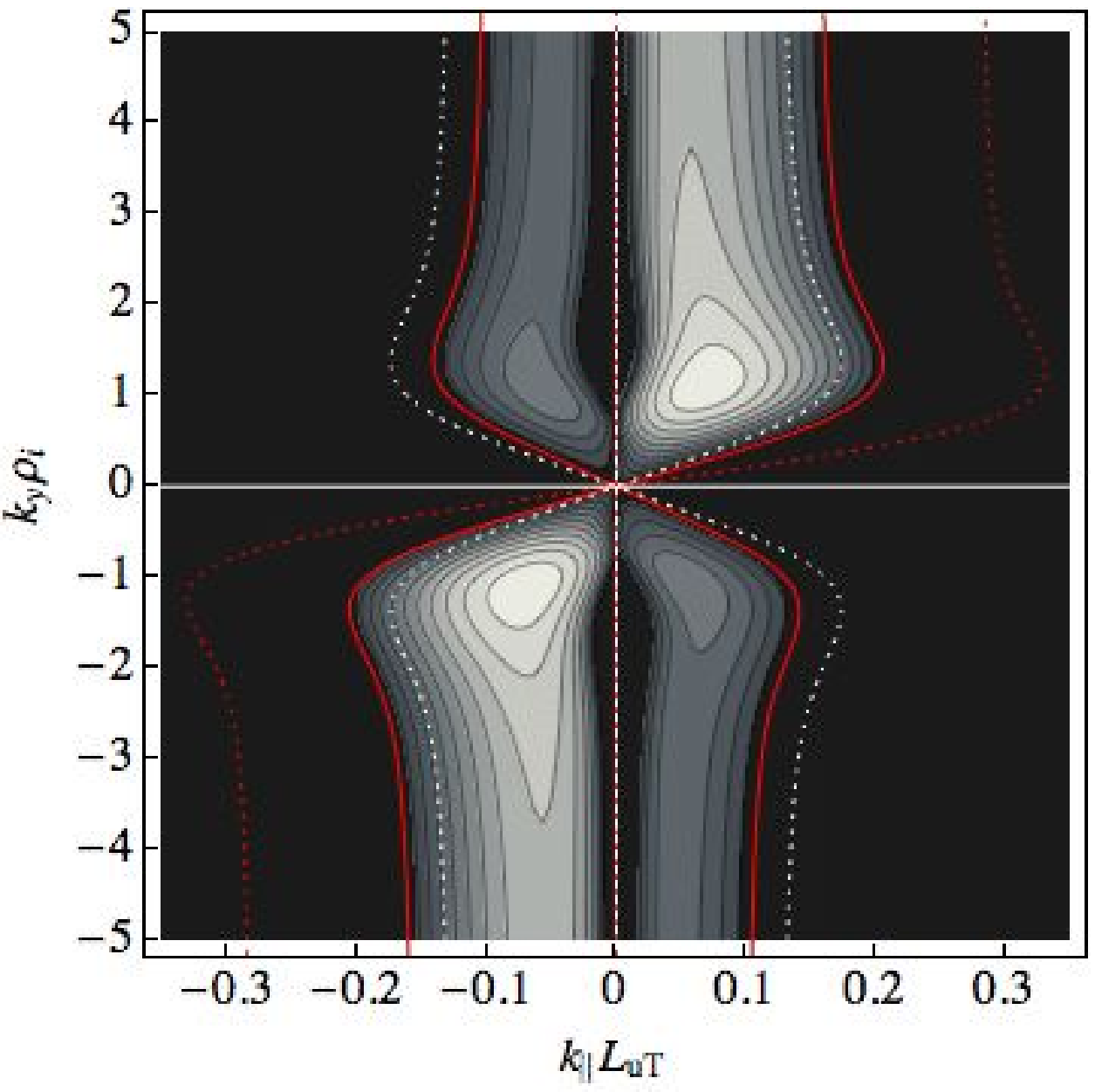,height=3in}}
\caption{\label{fig:itg} The ITG-PVG growth rate 
$\gamma/\sqrt{(qS/\ee)^2 + (\vthi/\LT)^2}$ vs.\ $\kpar\LuT$ and $k_y\rhoi$, 
where $\LuT=\vthi/\sqrt{(qS/\ee)^2+(\vth/\LT)^2}$. In this plot,  
$\eta_u=(\vthi/\LT)/(qS/\ee)=5$, $\eta_i=5$ and $\tau/Z=1$. 
Only positive values are plotted, black means $\gamma<0$.
The red curves show the stability boundary \eqref{eq:marg_gen}, 
the red dotted curves the pure-PVG ($\eta_u\to0$) stability boundary \eqref{eq:marg}
and the white dotted curve the pure-ITG ($\eta_u\to\infty$) stability 
boundary \eqref{eq:marg_itg}. See \figref{fig:itgkx} for cuts along the 
$k_y\rhoi$ axis for some representative values of $\kpar\LuT$.}
\end{figure}

\paragraph{Solutions.} It is not hard to show that the solutions of the ITG-PVG 
dispersion relation \eqref{eq:disp_rln_itg} can be expressed 
in the general form, which is a direct generalisation of 
\eqref{eq:gamma_gen_pvg}: 
\beq
\omega = \sqrt{(qS/\ee)^2 + (\vthi/\LT)^2}\,
f\lt(k_y\rhoi,\kpar\LuT;\eta_u,\eta_i\rt).
\label{eq:omega_gen}
\eeq
The growth rate $\gamma=\Im\,\omega$, obtained numerically for a typical situation 
with finite $\eta_u$, is shown in \figref{fig:itg}, as are the marginal stability 
thresholds \eqref{eq:marg_gen}, \eqref{eq:marg_itg} and \eqref{eq:marg}. 
Note that in general the ITG-PVG mode also has a real frequency.  
Note also that the solutions of the general ITG-PVG dispersion relation 
retain the PVG-mode property of independence of $k_y$ at short 
perpendicular wavelengths. This is easily demonstrated analytically  
by taking the limit $k_y\rhoi\gg1$, $\kpar\LuT$ in \eqref{eq:disp_rln_itg} 
--- all factors of $k_y\rhoi$ then cancel on both sides of the equation. 

\paragraph{Growth at long parallel wavelengths.} 
To assist physical intuition and 
some of the forthcoming discussion, it is useful to consider 
the ITG-PVG dispersion relation in the ``fluid'' limit, 
$\zz\gg1$ (case of long parallel wavelengths). Expanding 
$\Zfn(\zz)= - 1/\zz - 1/2\zz^3 - 3/4\zz^5 + \dots$, we recast 
\eqref{eq:disp_rln_itg} as a cubic equation
\bea
\nonumber
2\lt[\frac{1+\tau/Z}{\Gamma_0(\lambda)}-1\rt]\zz^3 
&+& 2\eta_i\zn\lt[1 - \Lambda(\lambda) - \frac{1}{\eta_i}\rt]\zz^2
+ \lt(\frac{q}{\ee}\,\zS-1\rt)\zz\\
&-&\eta_i\zn\lt[\Lambda(\lambda) + \frac{1}{\eta_i}\rt] = 0. 
\label{eq:cubic}
\eea
If we also consider long perpendicular wavelengths, $\lambda\to0$,
and assume $\eta_i\gg1$, we get
\beq
\frac{2\tau}{Z}\,\zz^3 + \lt(\frac{q}{\ee}\,\zS-1\rt)\zz - \eta_i\zn = 0.
\eeq 
The long-wavelength PVG instability \eqref{eq:sound_wave} is recovered 
as a balance of the first two terms (in the limit $\eta_u\ll1$); 
the fluid version of the ITG instability obtains from the balance of the 
first and third terms (when $\eta_u\gg1$):
\beq
\zz^3 \approx \frac{Z}{2\tau}\,\eta_i\zn
\quad\Rightarrow\quad
\omega \approx (\kpar^2 c_s^2 \eta_i\omega_*)^{1/3}
\label{eq:itg_fluid}
\eeq
(three roots, one real, two complex, of which one unstable). 

\begin{figure}[t]
\centerline{
\epsfig{file=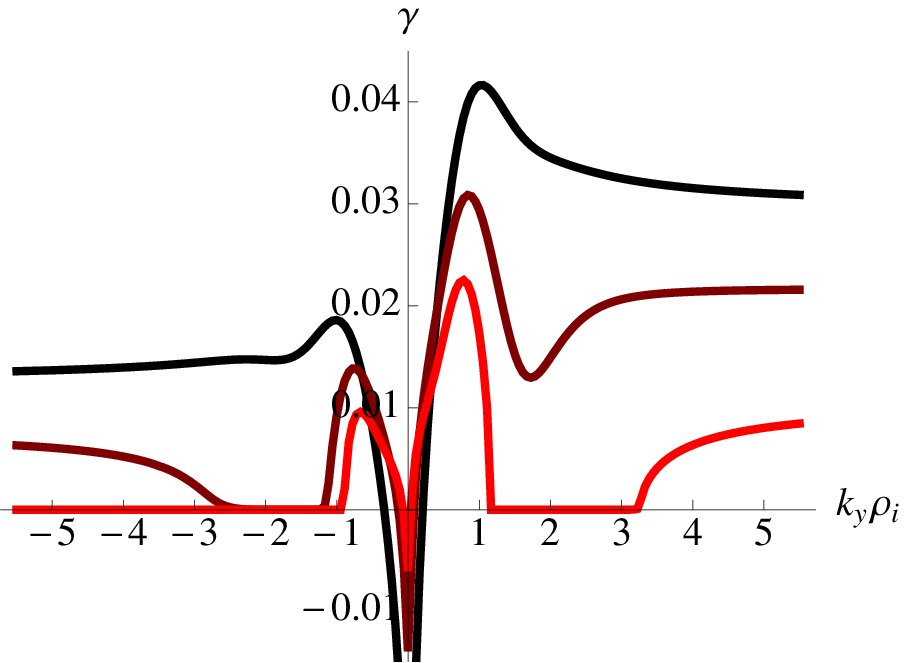,height=2in}
\hskip0.1in
\epsfig{file=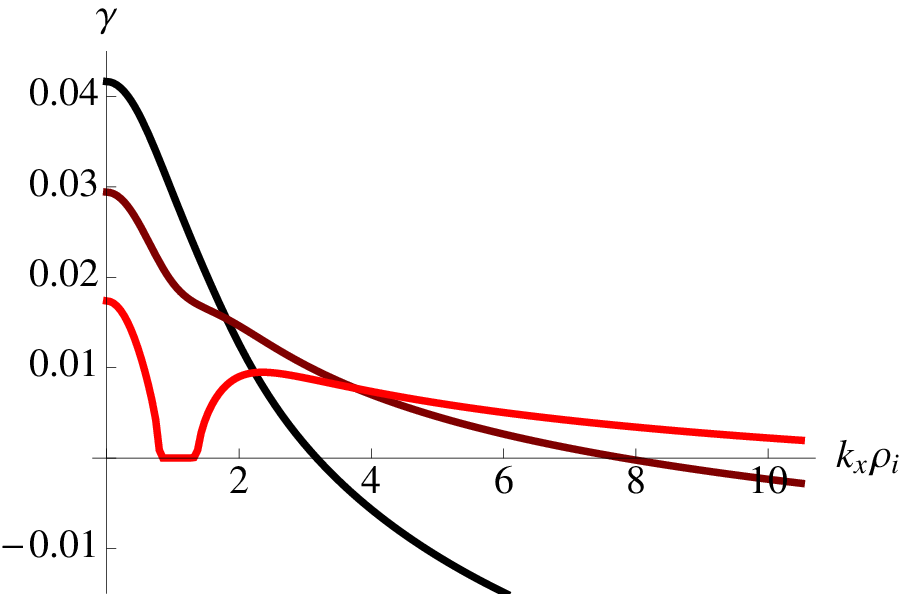,height=2in}
}
\caption{\label{fig:itgkx} {\em Left panel:} 
The ITG-PVG growth rate, calculated numerically from \eqref{eq:disp_rln_itg}
for the same parameters and  
normalised in the same way as in \figref{fig:itg}, 
vs.\ $k_y\rhoi$. The three curves are for 
$\kpar\LuT=0.01$ (red), $0.02$ (brown) and $0.05$ (black).
{\em Right panel:} Same, but vs.\ $k_x\rhoi$.}
\end{figure}

Analysing the long-wavelength dispersion relation in great detail 
is not a useful exercise because it turns out to be a very poor quantitative 
approximation to \eqref{eq:disp_rln_itg} in most parameter regimes. 
It does, however, make transparent how the ITG and PVG modes coexist. 
It also helps understand qualitatively what is perhaps a somewhat obscure 
property of the ITG (and ITG-PVG) mode: if the parallel wavenumber is held 
fixed and low ($\kpar\LuT\ll1$), the mode is unstable at long perpendicular 
wavelengths ($\lambda\ll1$), becomes stabilised at finite $\lambda$, and 
then reignites at larger $\lambda$. This is because the ``fluid'' ITG solution 
\eqref{eq:itg_fluid} depends on the cancellation $\Lambda(\lambda\to0)\approx1$ 
(and the limit $\eta_i\gg1$) in the second term of \eqref{eq:cubic}. As $\lambda$ 
increases, this second term becomes larger than the first and the mode is stabilised
(or, more precisely, it still has an exponentially small growth rate 
originating from the $i\sqrt{\pi}\,e^{-\zz^2}$ term in $\Zfn(\zz)$, neglected 
in the derivation of \eqref{eq:cubic} --- this comes from the Landau pole 
and is not recoverable in fluid theories). The instability is rekindled 
at larger $\lambda$ because the coefficient in the first (cubic) term in 
\eqref{eq:cubic} grows ($\sim\sqrt{\lambda}$ as $\lambda\to\infty$) and 
comes back into play (although the approximation $\zz\gg1$ breaks down 
simultaneously). 

This behaviour, calculated from the full dispersion relation 
\eqref{eq:disp_rln_itg}, is illustrated in \figref{fig:itg}. 
In the right panel of this figure, we show the growth rate dependence 
on $k_x\rhoi$, the dependence on which we have restored by 
setting $\lambda=(k_x^2+k_y^2)\rhoi^2/2$ (similarly to what was done 
at the end of \apref{app:pvg}). 
The attenuation and subsequent resumption of growth as the perpendicular 
wavenumber grows is the origin of similar behaviour, but as a function 
of time, seen in the long-time limit of the ITG-PVG transient growth 
in \secref{sec:incl_itg} and \apref{app:longt} --- because the effect of 
the perpendicular shear is to increase gradually the instantaneous $k_x$ 
of the mode. At long parallel wavelengths, as time increases, the mode 
first grows (when $St\ll1$), then slows down for a while, then (at $St\gg1$) 
the growth is resumed for another period and finally extinguished at $t=t_0$.  

\section{Transient growth with finite shear}
\label{app:longt}

Here we provide some asymptotic considerations that back up the behaviour 
of the time-dependent growth rates and frequencies described in \secref{sec:finiteS}. 

As in \secref{sec:weakS}, when $t\ll t_0$, we may expand the time-dependent dispersion 
relation \eqref{eq:itgt} assuming large frequency $\tz\gg1$, which 
is mostly real ($\tg\equiv\Im\,\tz\ll\Re\,\tz$): 
\beq
\fl
\lt[1 + \sigma\sqrt{1+ \lt(1-\frac{1}{\eta_i}\rt)\chi^2}\rt]
\frac{t}{t_0} \approx
-\lt(1-\frac{2}{\eta_i}\rt)\frac{\chi}{2\tz} - \frac{1}{\tz^2}
+ i\sqrt{\pi}\lt(2-\chi\tz\rt)\tz\,e^{-\tz^2}.
\label{eq:growth_itg}
\eeq
The two terms on the right-hand side that involve $\chi$ are the ITG 
terms and the other two are the PVG terms. When the latter or the former 
predominate, we recover the limiting cases treated in \secref{sec:weakS} 
and \secref{sec:longt}, respectively (in the case of strong PVG-driven 
growth the resonant term should be dropped as the growth rate is not small). 
The transition between the two regimes can be understood essentially by 
calculating finite-$\chi$ corrections. 

\paragraph{Strong ITG, weak PVG.}
Let us first consider the case where $\chi\tz\gg1$ but $\chi\sim 1$. 
Then the two PVG terms on the right-hand side of \eqref{eq:growth_itg} 
can be dropped and we obtain the following solution
\beq
\tz \approx -\frac{(1-2/\eta_i)\chi}{1 + \sigma\sqrt{1+ (1-1/\eta_i)\chi^2}}
\frac{t_0}{2t},\qquad
\tg \approx \frac{2\sqrt{\pi}}{1-2/\eta_i}\,\tz^4\,e^{-\tz^2}, 
\label{eq:sln_itg}
\eeq
a straightforward generalisation of \eqref{eq:growth_weakS}, to which 
it reduces when $\chi\to\infty$ and $\eta_i\to\infty$ (there is, of course, no 
problem retaining finite $\eta_i$, it was assumed large in \secref{sec:weakS} 
merely to make the algebra more compact). The exponentially small growth 
rate $\tg$ stops being exponentially small and takes off to order-unity 
values when the frequency $\tz$ stops being large and becomes $\tz\sim1$. 
It is evident from \eqref{eq:sln_itg} that this happens at earlier 
times for smaller $\chi$. Hence the behaviour of the growth rates 
and frequencies shown in \figref{fig:gamma_itg}: the peak growth 
moves into the past (smaller $t/t_0$) as $\chi$ decreases. 
The effect of finite $\chi$ is only noticeable when $\chi^2\sim1$; 
at larger values, the frequency is basically independent 
of $\chi$ and so the weak-shear limit ($\chi\to\infty$; \secref{sec:weakS}) applies.  

\begin{figure}[t]
\centerline{\epsfig{file=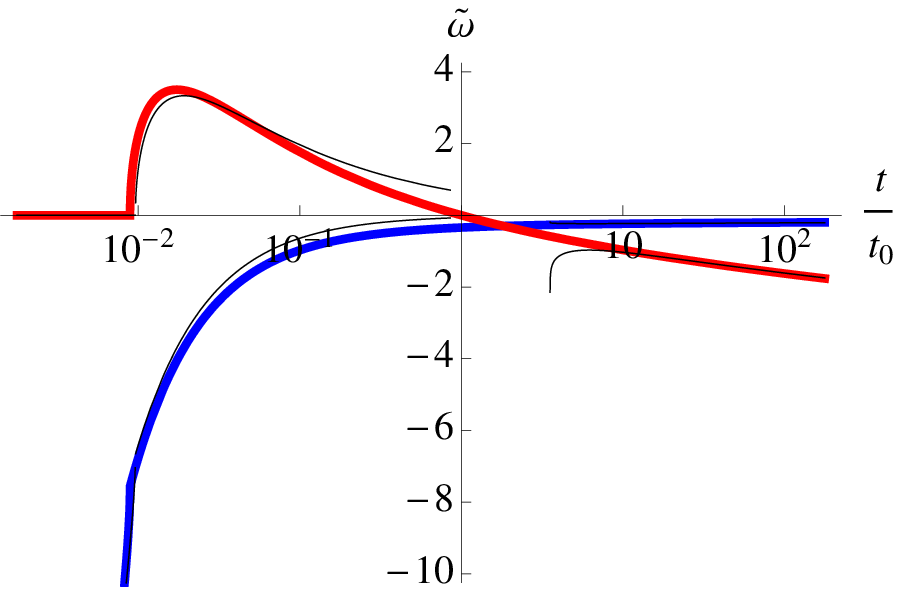,width=3.5in}}
\caption{\label{fig:gamma_chi1} Effective normalised growth rate $\Im\,\tz(t/t_0)$ (red, top) 
and frequency $\Re\,\tz(t/t_0)$ (blue, bottom) 
obtained via numerical solution of \eqref{eq:itgt} with 
$\eta_i=5$ and $\chi=1$. This is one of the cases already shown in \figref{fig:gamma_itg}, 
replotted here to compare with the analytic solutions derived in \apref{app:longt}: 
the black (thin) lines show the (combined) growth asymptotics 
\eqref{eq:omega_itg} and \eqref{eq:omega_approx}  
for $t/t_0\ll0$ (the growth rate \eqref{eq:gamma_approx} is so small 
that it is zero for all practical purposes) and the decay 
asymptotic \eqref{eq:decay_itg_precise} for $t/t_0\gg1$.} 
\end{figure}

\paragraph{Strong PVG, weak ITG.} Let us now restore the PVG terms
in \eqref{eq:growth_itg}. This becomes necessary when $\chi\tz\sim1$
or smaller. 
In this limit and in neglect of the exponentially 
small resonant term, the solution of~\eqref{eq:growth_itg}~is 
\beq
\tz \approx -\frac{\chi}{8}\lt(1-\frac{2}{\eta_i}\rt)\frac{\tpvg}{t} 
\pm i \sqrt{\frac{\tpvg}{2t}
-\lt[\frac{\chi}{8}\lt(1-\frac{2}{\eta_i}\rt)\frac{\tpvg}{t}\rt]^2}, 
\label{eq:omega_itg}
\eeq
where, to make the formula more compact, 
we have used the first expression in \eqref{eq:chi} to express 
the left-hand side of \eqref{eq:growth_itg} in terms of $\tpvg$. 
Here we are assuming $(q/\ee)\zS > 1$, so $\sigma=1$, $\tpvg>0$ and $\chi>0$.

The ``$+$'' branch in \eqref{eq:omega_itg} 
is a growing mode and is a direct generalisation 
of \eqref{eq:growth}. The difference with the pure-PVG case is that 
growth does not extend indefinitely into the past, but requires
\beq
\frac{t}{\tpvg} > \frac{\chi^2}{32}\lt(1-\frac{2}{\eta_i}\rt)^2.
\eeq
At longer times, the growth rates increases, reaches its maximum value 
\beq
\tg_\mathrm{max} = \frac{2}{\chi(1-2/\eta_i)}
\quad\mathrm{at}\quad 
\frac{t}{\tpvg} = \frac{\chi^2}{16}\lt(1-\frac{2}{\eta_i}\rt)^2,
\eeq
and then asymptotes to the pure-PVG case \eqref{eq:growth}.
At shorter times, the frequency \eqref{eq:omega_itg} becomes purely 
real (asymptoting eventually to $\tz$ given by \eqref{eq:sln_itg}). 
To get the growth rate in this regime, we must restore the resonant 
term in \eqref{eq:growth_itg}, which gives an exponentially small 
growth rate. The (growing) solution now is 
\bea
\label{eq:omega_approx}
\tz &\approx& -\frac{\chi}{8}\lt(1-\frac{2}{\eta_i}\rt)\frac{\tpvg}{t} 
- \sqrt{\lt[\frac{\chi}{8}\lt(1-\frac{2}{\eta_i}\rt)\frac{\tpvg}{t}\rt]^2
-\frac{\tpvg}{2t}},\\ 
\tg &\approx& \frac{\sqrt{\pi}(2+\chi|\tz|)|\tz|^3 e^{-\tz^2}}{\sqrt{\lt[\chi(1-2/\eta_i)\tpvg/8t\rt]^2-t_0/2t}}.
\label{eq:gamma_approx}
\eea 
\Figref{fig:gamma_chi1} shows the numerical solution of \eqref{eq:itgt} together with 
the asymptotics \eqref{eq:omega_itg}, \eqref{eq:omega_approx} and \eqref{eq:gamma_approx},
all calculated for $\chi=1$ and $\eta_i=5$. Even though the asymptotics are only technically 
valid for $\chi\lesssim1/\tz\ll1$, we see that they in fact work rather well 
even for moderate finite $\chi$ (this is due to the accumulation of 
numerically small prefactors that multiply $\chi$ in the above expressions). 
For values of $\chi$ significantly larger 
than unity, this agreement breaks down and one has to use \eqref{eq:sln_itg}.\\ 

Thus, we have learned that both PVG- and ITG-dominated dominated 
transient growth is exponentially weak at very short times, 
then increases to large or finite values before slackening again 
and turning to decay at $t=t_0$. At what time the transition 
happens and how large $\tg$ can get depends on $\chi$. 
Recalling the definition of $\chi$ (see \eqref{eq:chi}), we note that 
larger values of $\chi$ are achieved for stronger shear (i.e., $\eS$ larger 
compared to $q/\ee$), larger $k_y\rhoi$ or smaller $\kpar\vth/S$ 
(i.e., larger $\zS$). Thus, the transition between the ITG and PVG 
regimes happens non-uniformly in the wavenumber space 
(as, indeed, is evident in the middle panel of \figref{fig:N}). 

We remind the reader that all of the above 
is only valid in the long-time limit, $\zS t\gg1, k_y\rhoi$ 
(see the beginning of \secref{sec:longt}), which means that 
for some wavenumbers and/or values of $\eS$ and $q/\ee$, 
the transition from exponentially small to finite growth rate 
may be superseded by the transition from the short- to long-time 
limit, i.e., the initial ITG-PVG instability may not have 
time to peter out (due to increase in $\kperp\rhoi$ caused by 
shearing; see \apref{app:itg}) 
before being rekindled again by the  transient growth. 

\paragraph{Long-time decay.} 

Finally, to complete our treatment of the general ITG-PVG case, 
let us consider the eventual decay of the fluctuations. 
This is done in exactly the same way as in \secref{sec:decay}. 
In the limit $t\gg t_0$, we assume again that $\tz=i\tg$, where 
$\tg$ is large, negative and mostly real. Then \eqref{eq:itgt} 
is approximated by
\beq
\fl
\lt[1 +\sigma\sqrt{1+ \lt(1-\frac{1}{\eta_i}\rt)\chi^2}\rt]
\frac{t}{t_0} \approx
2\sqrt{\pi}\lt(2+i\chi|\tg|\rt)|\tg|\,e^{\tg^2}
\quad\Rightarrow\quad
\tg\approx -\sqrt{\ln\frac{t}{t_0}},
\label{eq:decay_itg}
\eeq 
so, like in \eqref{eq:decay}, we have a root-log law, i.e., the 
decay of the modes with time is just barely super-exponential.  
As before, \eqref{eq:decay_itg} is a quantitatively poor approximation 
except at ridiculously long times and a better one can be obtained by 
retaining corrections. This way we also obtain the (decaying with time) 
real frequency. The result is
\beq
\fl
\tg\approx - \sqrt{\ln\lt[\frac{1+\sqrt{1+(1-1/\eta_i)\chi^2}}{\sqrt{4+\chi^2\ln(t/(2\sqrt{\pi}\,t_0)}}
\frac{t/(2\sqrt{\pi}\,t_0)}{\sqrt{\ln(t/(2\sqrt{\pi}\,t_0)}}\rt]},\qquad
\tz\approx - \frac{\arctan(\chi|\tg|/2)}{2|\tg|}. 
\label{eq:decay_itg_precise}
\eeq
For $\chi\ll1$, we recover the pure-PVG result \eqref{eq:decay_precise}. 
For $\chi\gg1$ and $\eta_i\gg1$ (the weak-shear limit), 
these asymptotics are shown in \figref{fig:weakS}  
and for $\chi=1$ and $\eta_i=5$ in \figref{fig:gamma_chi1}.

\end{document}